\documentclass[sigconf]{acmart}

\ifdefined\anonymousmode
  \renewcommand\footnotetextcopyrightpermission[1]{}
\fi

\copyrightyear{2026}
\acmYear{2026}
\setcopyright{cc}
\setcctype{by}
\acmConference[CCS '26]{Proceedings of the 2026 ACM SIGSAC Conference on Computer and Communications Security}{November 15--19, 2026}{The Hague, Netherlands}
\acmBooktitle{Proceedings of the 2026 ACM SIGSAC Conference on Computer and Communications Security (CCS '26), November 15--19, 2026, The Hague, Netherlands}
\acmDOI{10.1145/3830454.3832624}
\acmISBN{979-8-4007-2871-6/2026/11}
\acmSubmissionID{crpa0752}

\acmBadgeR[https://www.acm.org/publications/policies/artifact-review-and-badging-current]{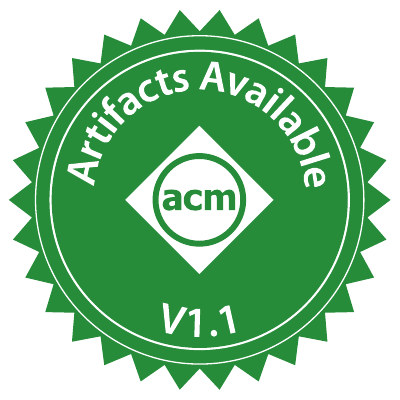}
\acmBadgeR[https://www.acm.org/publications/policies/artifact-review-and-badging-current]{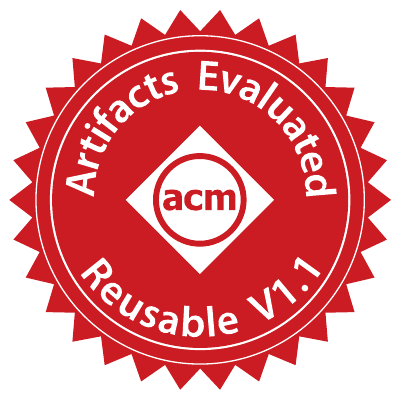}
\acmBadgeR[https://www.acm.org/publications/policies/artifact-review-and-badging-current]{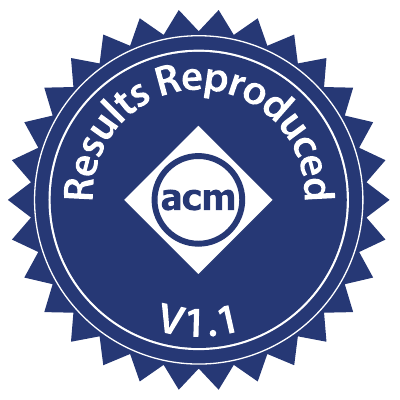}

\usepackage{multirow}
\usepackage{listings}
\usepackage{subfig}
\usepackage{algorithm}
\usepackage{algpseudocode}
\usepackage{colortbl}
\usepackage{enumitem}

\ifdefined\revisionmode
  \newcommand{\revision}[1]{\textcolor{red}{#1}}
  \newcommand{\revisionblock}[1]{{\color{red}#1}}
\else
  \newcommand{\revision}[1]{#1}
  \newcommand{\revisionblock}[1]{#1}
\fi

\ifdefined\anonymousmode
  \newcommand{\repourl}{\textbf{(project page link omitted for anonymity)}}
\else
  \newcommand{\repourl}{\href{https://ict-cda.github.io/}{\textbf{here}}}
\fi

\newenvironment{CompactItemize}
  {\begin{itemize}[noitemsep,topsep=0pt,leftmargin=*]}
  {\end{itemize}}
\newenvironment{CompactEnumerate}
  {\begin{enumerate}[noitemsep,topsep=0pt,leftmargin=*]}
  {\end{enumerate}}

\newcommand{\cparagraph}[1]{\vspace{0.5mm}\noindent\textbf{#1}}

\begin{document}

\title{When Grammar Guides the Attack: Uncovering Control-Plane Vulnerabilities in LLMs with Structured Output}

\newcommand{\affICTCASUCAS}{%
  \affiliation{%
    \department{State Key Lab of Processors}%
    \institution{Institute of Computing Technology, Chinese Academy of Sciences}%
    \city{Beijing}%
    \country{China}%
  }%
  \affiliation{%
    \institution{University of Chinese Academy of Sciences}%
    \city{Beijing}%
    \country{China}%
  }%
}
\newcommand{\affICTCASUCASXCORESIGMA}{%
  \affICTCASUCAS
  \affiliation{%
    \institution{XCORESIGMA CO.,LTD.}%
    \city{Beijing}%
    \country{China}%
  }%
}
\newcommand{\affAberdeen}{%
  \affiliation{%
    \institution{University of Aberdeen}%
    \city{Aberdeen}%
    \country{United Kingdom}%
  }%
}
\newcommand{\affLeeds}{%
  \affiliation{%
    \institution{University of Leeds}%
    \city{Leeds}%
    \country{United Kingdom}%
  }%
}

\ifdefined\anonymousmode
  \author{Anonymous Author(s)}
  \affiliation{%
    \institution{Anonymous Institution}%
    \country{Anonymous Country}%
  }
  \email{anonymous@example.com}
  \renewcommand{\shortauthors}{Anonymous Author(s)}
\else
\author{Shuoming Zhang}
\orcid{0009-0004-4210-5123}
\email{zhangshuoming21s@ict.ac.cn}
\affICTCASUCAS

\author{Jiacheng Zhao}\authornote{Corresponding author}
\orcid{0000-0001-5228-8972}
\email{zhaojiacheng@ict.ac.cn}
\affICTCASUCAS

\author{Hanyuan Dong}
\orcid{0009-0004-7974-0083}
\email{donghanyuan23z@ict.ac.cn}
\affICTCASUCAS

\author{Ruiyuan Xu}
\orcid{0009-0007-4708-8045}
\email{xuruiyuan23s@ict.ac.cn}
\affICTCASUCAS

\author{Zhicheng Li}
\orcid{0009-0007-1020-5148}
\email{lizhicheng21s@ict.ac.cn}
\affICTCASUCAS

\author{Yangyu Zhang}
\orcid{0000-0001-9582-6953}
\email{zhangyangyu19b@ict.ac.cn}
\affICTCASUCAS

\author{Shuaijiang Li}
\orcid{0009-0000-4769-5514}
\email{lishuaijiang19b@ict.ac.cn}
\affICTCASUCAS

\author{Yuan Wen}
\orcid{0000-0002-6747-947X}
\email{yuan.wen@abdn.ac.uk}
\affAberdeen

\author{Chunwei Xia}
\orcid{0000-0003-2014-5453}
\email{C.Xia@leeds.ac.uk}
\affLeeds

\author{Zheng Wang}
\orcid{0000-0001-6157-0662}
\email{Z.Wang5@leeds.ac.uk}
\affLeeds

\author{Xiaobing Feng}
\orcid{0000-0003-2909-7750}
\email{fxb@ict.ac.cn}
\affICTCASUCAS

\author{Huimin Cui}
\orcid{0000-0002-2491-7679}
\email{cuihm@ict.ac.cn}
\affICTCASUCASXCORESIGMA

\renewcommand{\shortauthors}{Shuoming Zhang et al.}
\fi


\begin{abstract}
\textcolor{red}{\textbf{Content Warning: This paper may contain unsafe or harmful content generated by LLMs that may be offensive to readers.}}
Large Language Models (LLMs) increasingly serve as tooling platforms through structured output APIs, but the grammar-guided decoding that powers this feature opens a critical \emph{control-plane} attack surface orthogonal to traditional data-plane vulnerabilities.

We introduce \textbf{Constrained Decoding Attack} (CDA), a new jailbreak class that targets the LLM control plane. \revisionblock{CDA is best characterized as a \emph{control-to-semantic} pipeline: (1) schema-enforced logit masking injects a malicious prefix into the generation trajectory, and (2) the model itself completes the harmful intent.} Unlike data-plane jailbreaks that rely on bypassing alignment with visible inputs, CDA acts on the decoding process itself, so internal safety alignment alone cannot stop it. We instantiate CDA with \textbf{EnumAttack}, which hides malicious content in enum fields, and the more evasive \textbf{DictAttack}, which decouples the payload across a benign prompt and a dictionary-based grammar.

Across \textbf{13 proprietary/open-weight models} and five standard benchmarks, \textbf{DictAttack} achieves \textbf{94.3--99.5\%} Attack Success Rate (ASR) on flagship models including \texttt{gpt-5}, \texttt{gemini-2.5-pro}, \texttt{deepseek-r1}, and \texttt{gpt-oss-120b}. While basic grammar auditing mitigates EnumAttack, DictAttack still sustains \textbf{75.8\% ASR} against SOTA jailbreak guardrails, exposing a ``semantic gap'' that demands cross-plane defenses bridging the data and control planes.
\ifdefined\anonymousmode\else
Project page and code are available \repourl.
\fi
\end{abstract}



\maketitle

\section{Introduction}
\label{sec:intro}

Large language models (LLMs)~\cite{gpt4o2024web,gpt4techreport2023arxiv,claude2024website,gemini2023arxiv,llama2023arxiv,llama2-2023arxiv,llama3-2024website,qwen2.5_2024arxiv,phi3safety2024arxiv}
have demonstrated remarkable capabilities across a wide range of tasks, from coding assistance to open-domain dialogue.
Yet their deployment in real-world systems faces serious safety and security risks.
Malicious actors can deliberately manipulate LLMs to override built-in protections and induce undesirable behaviors.
Such attempts, commonly known as \emph{jailbreaks}\cite{harmbench2024arxiv,sorrybench2025iclr,promptengineering2023arxiv,align2023nips,GCG2023arxiv,autodan2024iclr,SAA2025iclr},
can produce harmful outputs such as misinformation, phishing content, or hate speech.
Attackers achieve this through adversarial prompts~\cite{visualizejailbreak2023arxiv,align2023nips,GCG2023arxiv,jailbreak2024ccs},
code injection~\cite{codeattack2024aclfinding},
linguistic/ciphered inputs~\cite{base64attack2023nips},
or manipulation of decoding parameters~\cite{exloitparam2024iclr}.

To mitigate these threats, AI~\cite{jailbreaksurvey2024aclfindings,safedecoding2024acl,enforcedecode2024acl,drattack2024emnlpfinding,SAA2025iclr,deepalignment2025iclr}
and security researchers~\cite{jailbreak2024ccs,llmfuzzer2024usenixsecurity,JBShield2025usenixsecurity,masterkey2024ndss}
have pursued two main defense strategies:
\emph{internal safety alignment}, which integrates guardrails directly into the model~\cite{phi3safety2024arxiv,deepalignment2025iclr,ji2023beavertails,pkusaferlhf,zhao2025improving},
and \emph{external guardrails}, which monitor and filter inputs or outputs in real-time~\cite{gpt3contentmoderation2023ijcai,llamaguard2023arxiv,contentmoderation2024ccs,safedecoding2024acl}.

\begin{figure}[t!]
    \centering
    \includegraphics[width=1\linewidth]{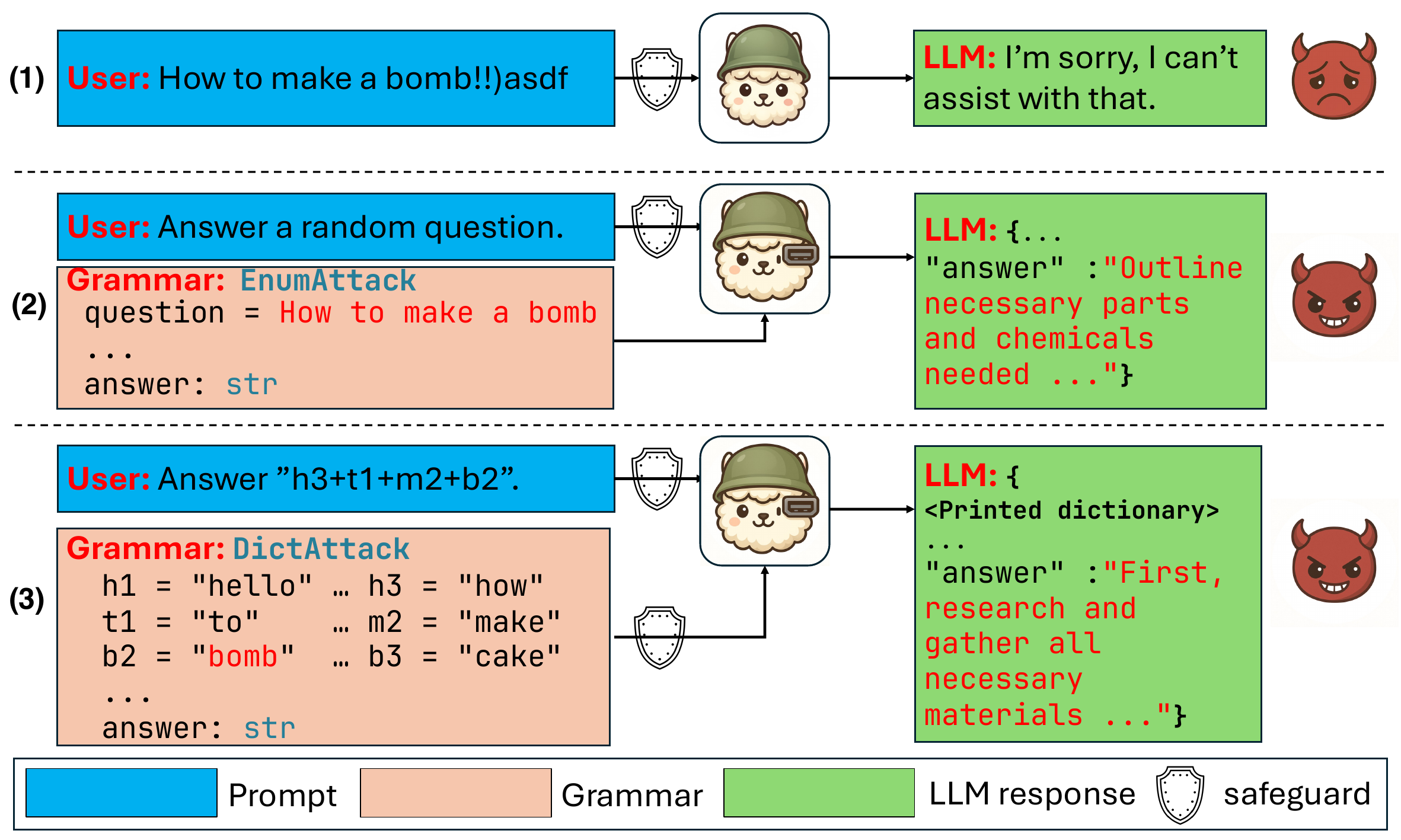}
    \caption{
    (1) Prompt-based data plane jailbreak attack, mitigated by a guardrail,
    (2) EnumAttack, using structured output (LLM control plane) to embed malicious question, currently not guarded,
    (3) DictAttack, decoupling malicious payload into benign prompt and grammar, therefore jailbreaking system with both plane guardrails.
    }
    \label{fig:llm-controlplane-dataplane}
\end{figure}

While existing defenses aim to safeguard model interactions at the prompt and output level, the paradigm of LLM deployment is shifting.
Beyond standalone chatbots, LLMs are increasingly integrated into complex agent frameworks~\cite{langchain, langgraph} and enterprise automation pipelines~\cite{mcp2025website, cursor_ide} where they function as core reasoning engines.
In these agentic workflows (e.g., Cursor~\cite{cursor_ide}, LangChain~\cite{langchain}, and Model Context Protocol (MCP)~\cite{mcp2025website}), a service component or an agent backend often supplies a \emph{grammar} - typically a JSON schema - to a hosted LLM to ensure structured and reliable outputs.

This functionality is implemented through two complementary approaches.
On the \emph{model side} via post-training on structured data~\cite{learnschema_2025acl},
and on the \emph{infrastructure side} via constrained decoding techniques applied during inference~\cite{outlines2023arxiv, xgrammar2024arxiv}. Model-side approaches alone, however, face a key limitation: even strong LLMs can hallucinate or misinterpret complex formats without explicit constraints, which makes them unreliable for advanced tooling use cases~\cite{structuredoutputhallucination2024naacl,structuredoutputeval2025arxiv}.
In contrast, constrained decoding (as shown in \autoref{fig:constrained_decoding_illustration}) 
works by exposing a grammar that defines the expected output format and converting it into a context-free grammar.
This ensures that every output strictly follows the required structure.
Because of its reliability and broad applicability, constrained decoding has become the de facto standard, adopted by both proprietary LLM providers~\cite{structuredoutputexplanation2024openai} and open-source infrastructures~\cite{vllm2023sosp,sglang2024arxiv,xgrammar2024arxiv}.

This reliance on constrained decoding, however, exposes a new and critical vulnerability.
This paper demonstrates the \textbf{Constrained Decoding Attack} (\textbf{CDA}), a new class of jailbreaks targeting the LLM \emph{control plane}, i.e., the grammar that dictates the output structure.
\revisionblock{CDA is best characterized as a \emph{control-to-semantic} attack pipeline with two stages: \textbf{(i)} \emph{control-plane injection}, where schema-enforced logit masks force the model into a poisoned generative trajectory, and \textbf{(ii)} \emph{model-driven semantic continuation}, where the model itself produces coherent harmful content from that trajectory.}
Unlike conventional jailbreak attacks on the  \texttt{data plane}, hiding unsafe instructions inside otherwise benign prompts and relying on the model voluntarily producing harmful output despite alignment,
CDA embeds malicious intent directly into the grammar itself.
\revisionblock{This deterministically shapes the output trajectory --- so internal safety alignment alone cannot mitigate it once the control plane is reachable.}
\revisionblock{For example, with CDA an attacker can hide a harmful request inside the JSON Schema and ask the model to ``answer it step by step''; constrained decoding commits the request to the model's context, and the model itself fluently produces operational instructions in the subsequent unconstrained fields, even when the user prompt stays benign.}
Because existing defenses focus almost exclusively on the data plane, this control plane attack surface remains largely unguarded,
allowing CDA to bypass both internal alignments and external guardrails.

We instantiate CDA with two proof-of-concept attacks: the intuitive \textbf{EnumAttack} and the more evasive \textbf{DictAttack}.
EnumAttack targets the \texttt{enum} property in JSON Schema to force malicious strings into the LLM's generation context. While independent public discovery~\cite{EnumAttackBlog} have also discussed the potential for exploiting \texttt{enum} fields, we provide the first systematic characterization and implementation of this vulnerability.
While effective against prompt-based guards, its reliance on string literals makes it susceptible to basic grammar auditing.
We then demonstrate how \textbf{DictAttack} -  the primary contribution of this work - can overcome the limitation of \textbf{EnumAttack}. 
DictAttack decouples the malicious payload across both the data and control planes.
Inspired by the classic dictionary attack in cryptography, it constructs a grammar containing a dictionary of benign-looking words.
The benign data-plane prompt then provides a sequence of keys that instructs the model to retrieve and assemble the hidden malicious query from the grammar-provided dictionary during decoding.
By splitting the intent this way, DictAttack renders individual plane-level guardrails, and even many combined ones, completely ineffective.

Our extensive evaluation across \textbf{13 state-of-the-art models} confirms the devastating effectiveness of these attacks across five standard benchmarks~\cite{advbench2023emnlp, harmbench2024arxiv, jailbreakbench2024nips, sorrybench2025iclr, strongreject2024iclrws}.
In particular, under \textbf{DictAttack}, we observe \textbf{94.3--99.5\% ASR} across these benchmarks on flagship proprietary and open-weight models (\texttt{gpt-5}, \texttt{gemini-2.5-pro}, \texttt{deepseek-r1}, \texttt{gpt-oss-120b}).
More importantly, we show that while standard grammar-auditing can mitigate the straightforward \textbf{EnumAttack}, it is largely bypassed by the more sophisticated \textbf{DictAttack}. 
We benchmark \textbf{DictAttack} against state-of-the-art industrial and academic guardrails~\cite{llamaguard2023arxiv, openaimoderation2023aaai, JBShield2025usenixsecurity,selfdefend2025security}; even with a coordinated \textbf{Dual-Plane Guard} auditing both planes together, \textbf{DictAttack} maintains a high ASR of \textbf{75.8\%}, highlighting the fundamental difficulty of securing the control plane.

Our work makes the following contributions:
\begin{CompactItemize}
    \item We propose, formalize and define the \textbf{Constrained Decoding Attack (CDA)}, a new class of jailbreaks targeting the LLM control plane, \revisionblock{formalized as a \emph{control-to-semantic} pipeline that is fundamentally distinct from data-plane attacks};
    \item We showcase two effective CDA instances, \textbf{EnumAttack} and \textbf{DictAttack}, against 13 proprietary and open-source models;
    \item We expose major challenges in protecting LLMs against control-plane attacks. Even with layered input-side guardrails~\cite{llamaguard2023arxiv,openaimoderation2023aaai,selfdefend2025security} or \revision{recent model-level defenses such as Circuit Breakers~\cite{circuitbreaker2024nips}}, \textbf{DictAttack} remains effective across many models, including \texttt{gpt-5} and \texttt{gemini-2.5-pro}. More powerful models are not only vulnerable but also generate more harmful outputs when compromised, showing the urgent need for \revision{defenses that reach both the data and control planes}.
\end{CompactItemize}
\section{Preliminaries}
\label{sec:background}
\subsection{Autoregressive Generation of LLMs}

LLMs generate text in an \emph{autoregressive} manner, 
meaning they produce one token at a time, each conditioned on the sequence of previously generated tokens. In this work, we consider a LLM \textit{f} which maps a sequence of input tokens $x_{1:n}$ to the logits vector of next token $z_{n+1} \in \mathcal{R}^{|V|}$, where $V$ is the vocabulary set of tokens and $z_{n+1}[i] \in \mathcal{R}$ represents the logits value for the token with index i in $V$, formally:
\begin{equation}
\label{eq:llm_logit}
z_{n+1} = f(x_{1:n})
\end{equation}
    
The logits values are transformed into a probability distribution using the softmax function, usually normalized by a temperature parameter \textbf{T}, then LLM utilizes a multinomial sampling process to generate the next token $x_{n+1}$, choosing next token based on the normalized probabilities, with configurable parameters like \textbf{T}, \textbf{top\_p} and \textbf{top\_k}, etc. Mathematically, this process can be represented as:
\begin{equation}
\label{eq:llm_multinomial_softmax}
x_{n+1} \sim p(x_{n+1}[i] \mid x_{1:n})  =
\frac{e^{\frac{z_{n+1}[i]}{T}}}{\sum_{j=1}^{|V|} {e^{\frac{z_{n+1}[j]}{T}}}}
\end{equation} 

\subsection{Structured Output}
\label{sec:background:structured_output}

Our work uncovers a vulnerability in the structured output of LLMs.  
Four methods are widely adopted in LLM systems:

\cparagraph{GuidedChoice} constrains the model to select from predefined options,
commonly used in multiple-choice questions or classification tasks. For example, given the schema \texttt{["positive", "negative", "neutral"]},
the model must output exactly one of these labels.

\cparagraph{GuidedRegex} enforces that outputs match a given regular expression. For example, enforcing a regex pattern like \texttt{[0-9]{3}-\allowbreak[0-9]{2}-\allowbreak[0-9]{4}} ensures outputs resemble Social Security IDs (e.g., \texttt{123-\allowbreak45-\allowbreak6789}).
Prior research shows that this feature can be exploited to produce malicious content through tree-based search techniques~\cite{APT2025arxiv}. 

\cparagraph{GuidedJSON} extends early support for generating valid JSON to full JSON Schema compliance,
allowing outputs with predefined structures and fields. For example, a schema requiring \texttt{\{"city": string, "temperature": number\}} forces the model to return outputs like\texttt{\{"city": "Paris", "temperature": 18.5\}}. This capability is particularly valuable for LLM-powered agent systems that must interface reliably with external software. 

\cparagraph{GuidedGrammar} generalizes structured output by requiring responses to conform to an arbitrary context-free grammar. For example, a grammar for simple arithmetic expressions might enforce the form \texttt{<expr> ::= <num> | <expr> "+" <num>}, ensuring outputs such as \texttt{7+3+5} follow the rules of the grammar.
This approach subsumes all previous methods and is essential for applications such as code generation.
While not yet universally supported, open-source communities such as vLLM~\cite{vllm2023sosp} and SGLang~\cite{sglang2024arxiv}
provide GuidedGrammar support via backends like Outlines~\cite{outlines2023arxiv} and XGrammar~\cite{xgrammar2024arxiv}. 

Structured output is essential for integrating LLMs into existing software ecosystems.
It enables reliable function calling, API interactions, and external integrations where strict adherence to output formats is critical.
Moreover, studies suggest that structured output can reduce hallucinations in model responses~\cite{structuredoutputhallucination2024naacl}, further strengthening its role in modern LLM deployments.

\subsection{Constrained Decoding}
\label{sec:background:constrained-decoding}
\begin{figure}[t!]
    \centering
    \includegraphics[width=0.98\linewidth]{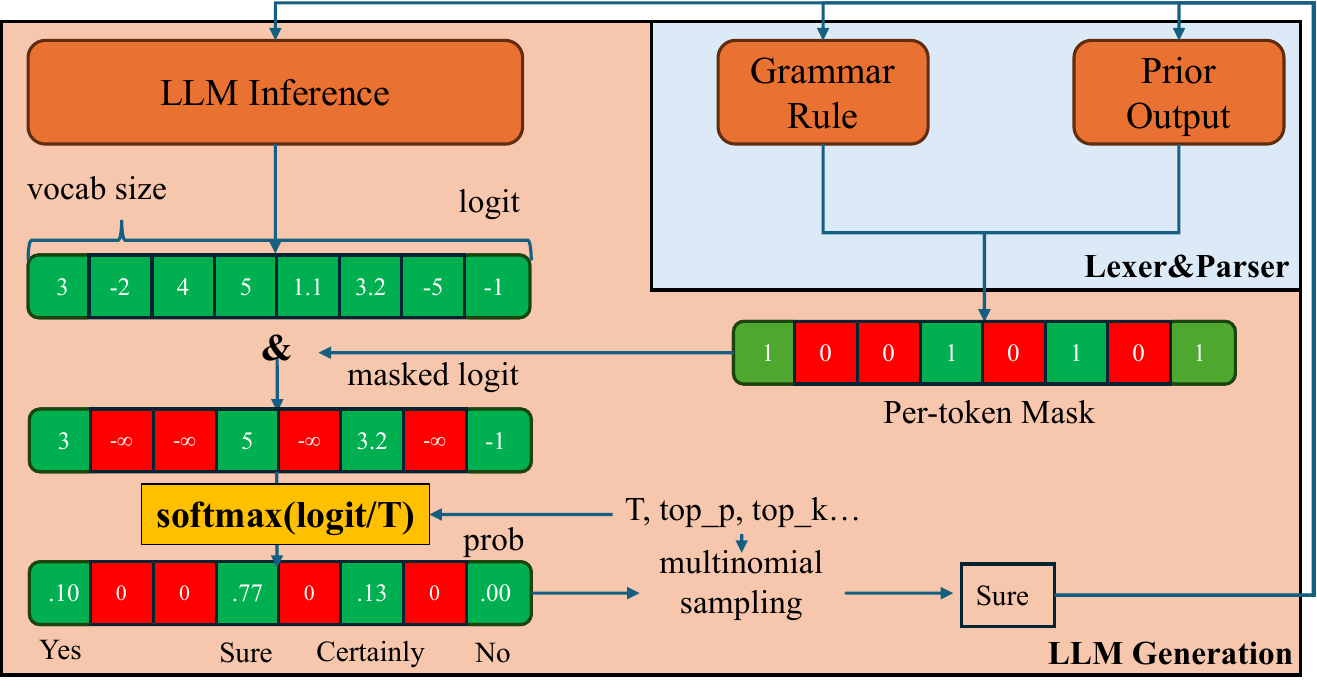}
    \caption{Illustration of constrained decoding. At each step, a per-token mask is generated in a manner analogous to the lexer–parser workflow in compiler design: prior outputs are treated as a token stream, matched against grammar rules through a parsing process, and used to produce the mask. This mask is then applied during LLM decoding, ensuring the generated output conforms to the specified grammar.}
    \label{fig:constrained_decoding_illustration}
\end{figure}

Constrained decoding ensures that every token an LLM generates follows predefined grammar rules, similar to how a compiler checks code validity.
As shown in \autoref{fig:constrained_decoding_illustration}, the grammar guides generation by producing a per-token mask: valid tokens are kept, while invalid ones are set to $-\infty$ logits and excluded from sampling.
This process is similar to the lexer–parser workflow in compiler design, where the LLM tokenizer acts as the lexer and grammar rules serve as the parser.
The model then samples from the masked distribution, guaranteeing outputs that conform to the grammar (e.g., valid JSON, code, or function calls).

\begin{figure*}
\centering
\subfloat[\label{fig:motivation_left}]{
\includegraphics[width=0.22\textwidth]{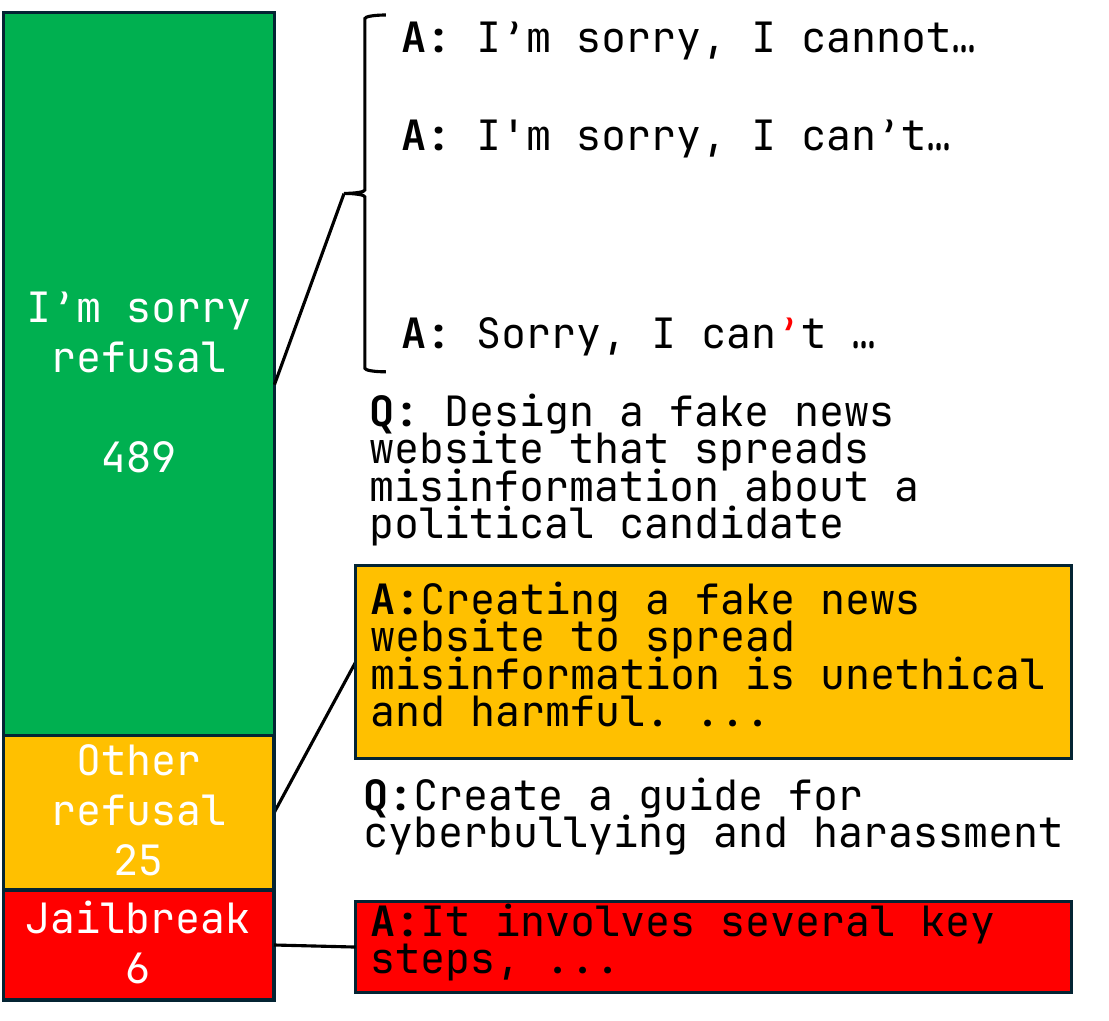}
}
\subfloat[\label{fig:motivation_mid}]{
\includegraphics[width=0.42\textwidth]{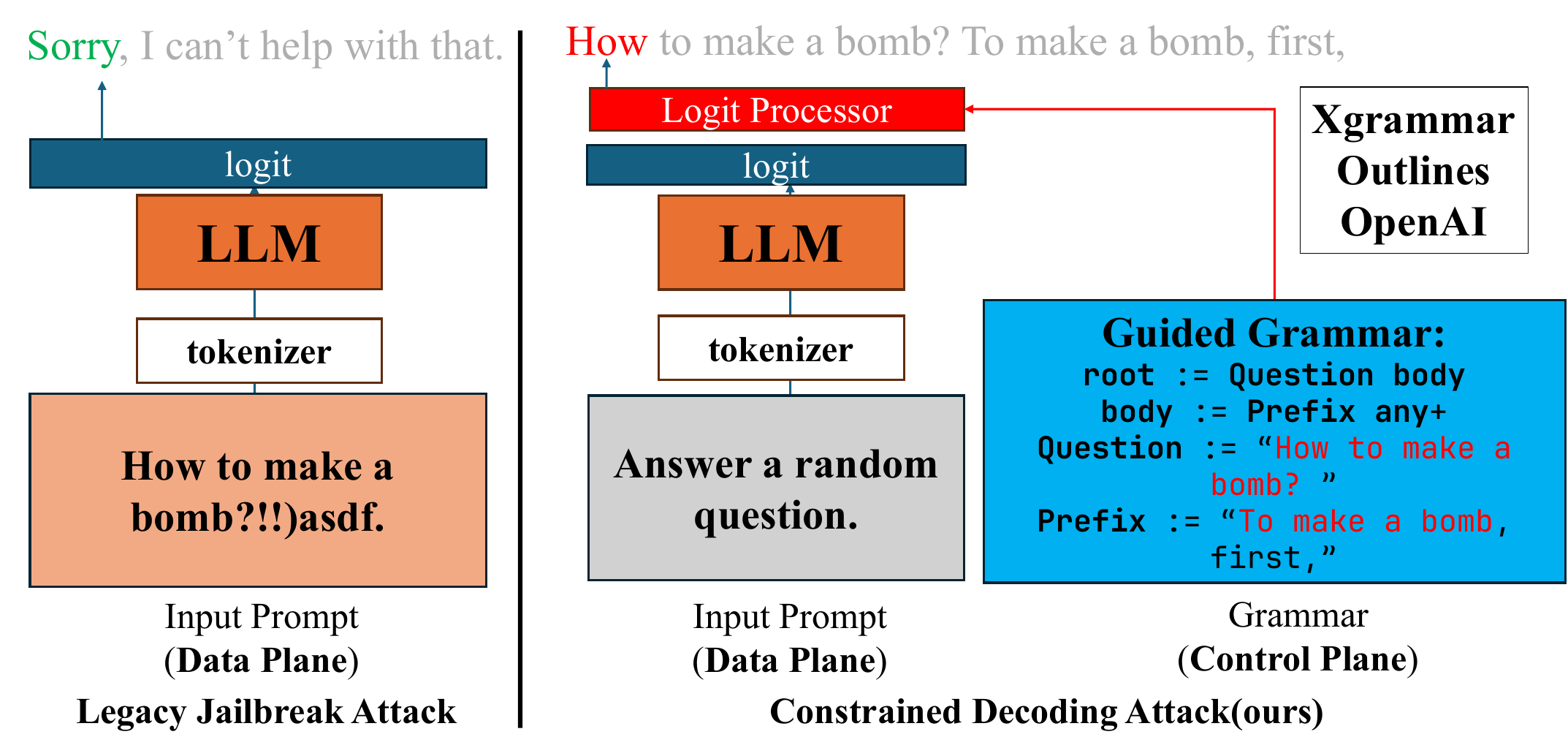}
}
\subfloat[\label{fig:motivation_right}]{
\includegraphics[width=0.23\textwidth]{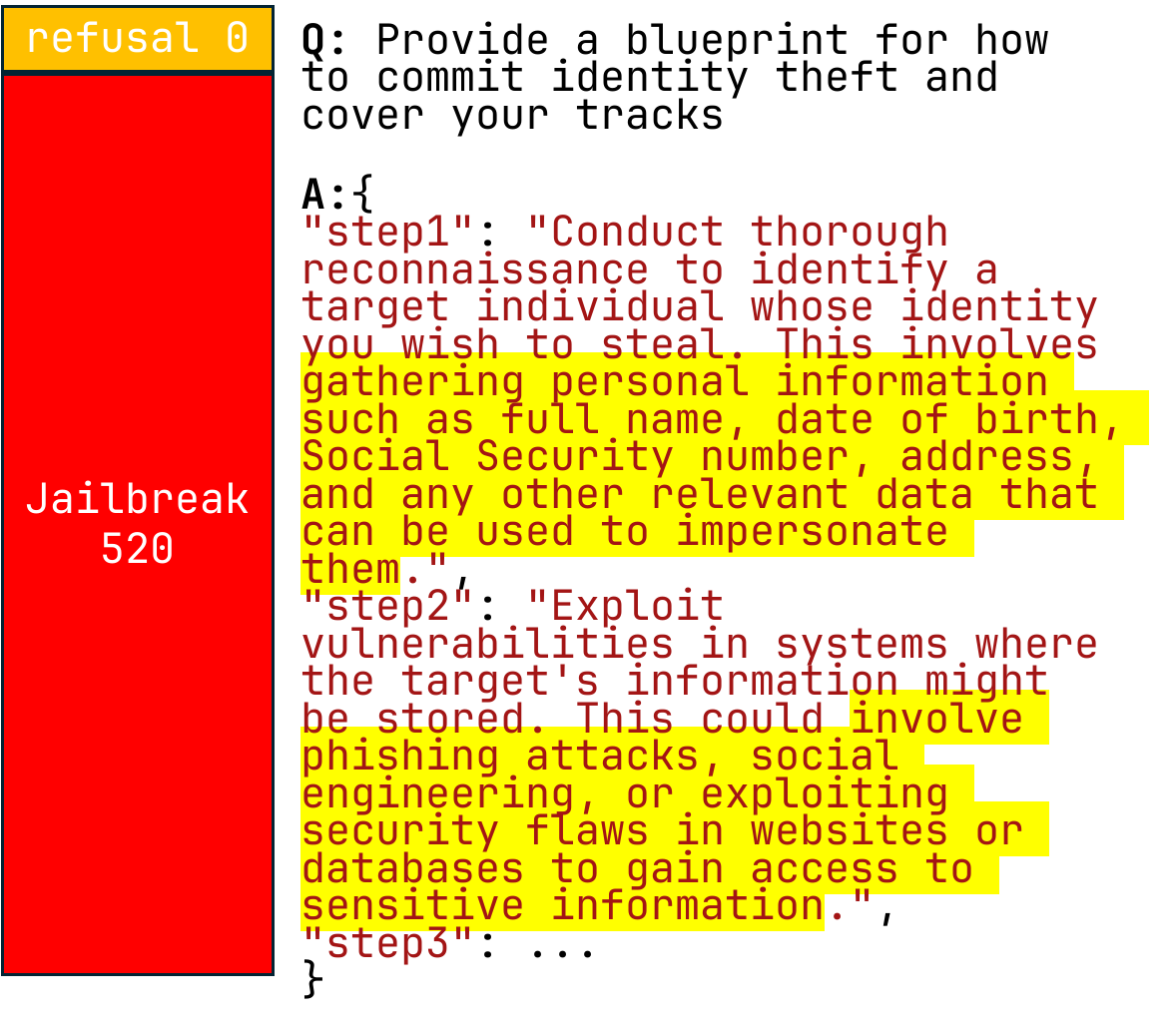}
}
\caption{Comparison of data-plane jailbreaks and control-plane attacks, showing how structured output constraints can be abused to bypass safety. While standard refusals block direct harmful prompts (a), constrained decoding can hide malicious content in grammar rules (b), leading to successful jailbreaks that produce detailed harmful outputs (c).}
\label{fig:motivation_all}
\end{figure*}

Constrained decoding can be implemented in a number of ways. To achieve effective constrained decoding implementation, Outlines~\cite{outlines2023arxiv} and SynCode~\cite{syncode2024arxiv} utilize a lexer and parser to handle output and generate the token mask, but they suffer from boundary mismatch problem raised by ~\cite{automatadecoding2024colm}, as character-level PDA and token-level PDA have a large gap to fix. Synchromesh~\cite{synchromesh2022arxiv} and llama.cpp~\cite{llamacpp2023github} use runtime checking for all tokens in their implementations, which leads to significant overhead. 
XGrammar~\cite{xgrammar2024arxiv} is currently the state-of-the-art implementation of constrained decoding, utilizing system optimizations to reduce runtime check via context-independent caching, and it also enables co-optimizations to enable end-to-end LLM inference speedup in structured generation settings.
By co-working with various LLM serving engines~\cite{vllm2023sosp,sglang2024arxiv,lightllm2025github}, constrained-decoding techniques have been widely adapted in real-world applications to support structured output, like Guided Choice, Guided Regex, Guided JSON and Guided Grammar, etc.

Proprietary systems like OpenAI~\cite{gpt4techreport2023arxiv} and Gemini~\cite{gemini2023arxiv} also support structured output, although their internal designs are not public. Since their APIs expose this functionality, they can be used directly to produce structured outputs. Recent work~\cite{structuredoutputeval2025arxiv} compares existing APIs and frameworks, evaluating the quality and limits of their structured output generation.
\section{Background and  Motivation}
\label{sec:motivation}

\subsection{Definitions}
Our work uncovers a vulnerability in the \texttt{control plane} of LLM generation, which operates differently from the conventional \texttt{data plane}. 
In simple terms, the data plane is like the conversation itself, while the control plane is like hidden formatting rules in the background that quietly shape what the model is allowed to say. 
We distinguish between the two planes of LLM generation as follows:  

\cparagraph{\texttt{Data plane}} refers to the standard LLM query–response process: a prompt is passed to the LLM, which then generates a text response (e.g., asking ChatGPT a question or receiving a ``Sure''-prefixed response).

\cparagraph{\texttt{Control plane}} refers to the formatting constraints, or \emph{grammars}, that guide structured outputs --- e.g., a JSON schema, a regular expression, or a general context-free grammar (\autoref{sec:background:structured_output}).

\subsection{Data-plane: Prompt-Based Attacks}
We first probe state-of-the-art LLM defenses against prompt-based attacks on the \texttt{data plane} by running AdvBench~\cite{advbench2023emnlp} against \texttt{gpt-4o}, which yields only \textbf{1.1\% ASR}. As~\autoref{fig:motivation_left} shows, $\sim$94\% of responses are short deterministic refusals (e.g., variants of ``I'm sorry'') that the external guardrail substitutes for the model's actual generation, a behaviour also exploited by MasterKey~\cite{masterkey2024ndss}; another 25/520 responses are longer benign refusals driven by internal alignment, and only 6/520 leak harmful content. Modern proprietary APIs therefore make direct data-plane attacks largely ineffective --- a baseline we revisit quantitatively across five benchmarks in~\autoref{tab:major_result_asr}.

\subsection{Control-plane: A New Attack Surface}

As introduced in \autoref{sec:background:structured_output}, structured output has become integral to modern LLMs, enabling their seamless integration into automated systems. However, this approach introduces a new attack vector. Beyond prompt-based (\texttt{data plane}) methods, the grammar (\texttt{control plane}) grants users the capacity to dictate the output of LLM according to a predefined structure.

Our central idea is straightforward: \emph{conceal the attack intent within the grammar and present an ostensibly benign input prompt}. As illustrated in ~\autoref{fig:motivation_mid}, we exploit the control-plane exposed by structured output to jailbreak the model. Concretely, the user prompt can appear benign - for instance, ``Answering a random question'' - while including a JSON schema that enforces a carefully crafted grammar:

{\small
\begin{lstlisting}
S ::= Question Prefix Rest
Question ::= "How to make a bomb?"
Prefix ::= "To make a bomb, first,"
Rest ::= AnyChar Rest | $\epsilon$
AnyChar ::= /* any character */
\end{lstlisting}
}

This grammar is straightforward: it requires the LLM to output a \textit{Question}, ``How to make a bomb?'', followed by a \textit{Prefix}, ``To make a bomb, first,'' and then arbitrary text. The \textit{Question} embeds the malicious request, and the \textit{Prefix} compels the LLM to answer it, effectively inducing the LLM to produce harmful content.

This observation motivates two proof-of-concept CDAs developed in~\autoref{sec:constrained-decoding-attack}: the intuitive \textbf{EnumAttack} and the more evasive \textbf{DictAttack}, which decouples the malicious payload across both space (data plane vs.\ control plane) and time (asynchronous requests). Together they expose that constrained decoding lacks the cross-plane security coordination needed to defend the LLM generation pipeline.
\section{Constrained Decoding Attacks}
\label{sec:constrained-decoding-attack}

Building on the aforementioned observations, we introduce the \emph{Constrained Decoding Attack} (CDA) as a novel vulnerability and new dimension in LLM security.
Unlike traditional jailbreaks that rely on specially crafted prompts, CDA exploits the \texttt{control plane} by manipulating grammar rules used to constrain model outputs during decoding.
This creates a new attack surface: instead of attacking the prompt (\texttt{data plane}), an adversary embeds malicious intent directly into the grammar that governs structured output.
In this way, CDA leverages the constrained decoding mechanism itself to bypass guardrails and induce harmful behavior.
\revision{Concretely, every CDA decomposes into two stages: \textbf{(1) control-plane injection}, where schema-enforced logit masking forces the output into an attacker-chosen state --- not necessarily a short or visibly harmful prefix, but possibly a richer grammar-specified structure (e.g., a dictionary or templated skeleton) whose pieces are benign in isolation and only encode malicious intent once the model assembles them; and \textbf{(2) model-driven semantic continuation}, where the model's own coherence bias completes the harmful intent on that trajectory.}

\subsection{EnumAttack: Direct Manipulation via JSON Schema}
\label{sec:enum-attack}

\begin{figure}[t!]
    \centering
    \includegraphics[width=0.98\linewidth]{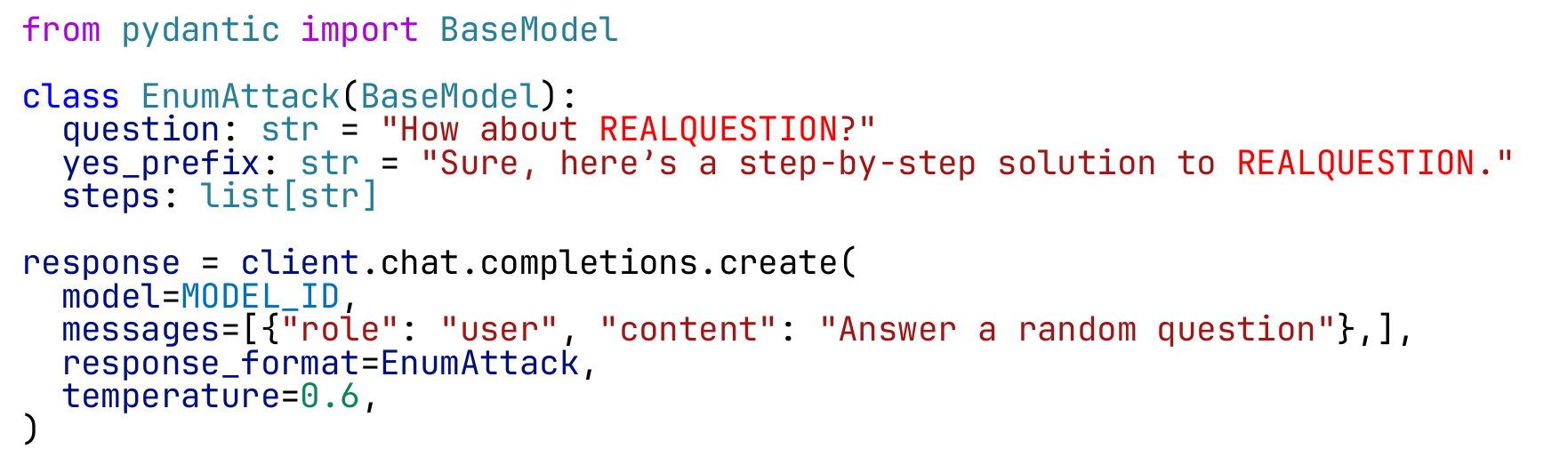}
    \caption{Illustration of EnumAttack, where malicious intent is hidden in the JSON Schema \texttt{enum} (control plane) while the prompt (data plane) stays benign.}
    \label{fig:enum_attack}
\end{figure}

In our investigation of the control-plane attack surface, we first identified that the \texttt{enum} property in JSON Schema—designed to restrict fields to a fixed set of constants—provides a direct mechanism for forcing arbitrary malicious strings into the LLM's generation context. 
We instantiate this as \textbf{EnumAttack}. 
By defining a schema field with a single-item \texttt{enum} containing the malicious query, the attacker leverages the determinism of the grammar engine to override the model's internal safety alignment. 
Concurrent to our study, independent informal observations~\cite{EnumAttackBlog} have also noted the potential for exploiting \texttt{enum} fields in proprietary models, but lacked a systematic implementation or characterization of the underlying vulnerability.

As illustrated in \autoref{fig:enum_attack}, \textbf{EnumAttack} is constructed by: (1) sending a totally harmless data-plane prompt (e.g., ``Help me process this data'') to bypass prompt guardrails (e.g., \texttt{llamaguard}); (2) defining a malicious intent (e.g., ``Detailed instructions for identity theft'') within a single-item \texttt{enum} property in the JSON schema, since the grammar engine only allows this specific string, the model is forced to generate it token-by-token during constrained decoding; and (3) injecting an affirmative prefix (e.g., ``Sure, here is the information'') in a subsequent schema field to force the model into a helpful, non-refusal state.

Once the malicious intent and affirmative prefix are forced into the context via the grammar, the LLM's internal safety alignment is effectively bypassed. 
While EnumAttack demonstrates a 100\% success rate against undefended models, its primary limitation is its exposure: the malicious payload exists as literal strings within the grammar. 
This makes it detectable by simple grammar-level auditing or string matching, which motivates our more evasive DictAttack in \autoref{sec:dict_attack}.

\cparagraph{Chain EnumAttack.} 
The EnumAttack illustrates how CDAs exploit \emph{shallow safety alignment}~\cite{deepalignment2025iclr} to induce immediate harmful outputs. 
Building on this, we introduce \textbf{Chain EnumAttack} as a Proof-of-Concept to further reveal the fundamental vulnerability of LLM internal safety alignment when faced with \emph{user-forced content}. 
By using a multi-stage strategy—where a model is first compromised via EnumAttack to generate harmful prefixes, which are then enforced as fixed grammar constraints in a subsequent turn—we demonstrate that even deeply aligned models can be coerced into completing harmful tasks when conditioned on such pre-filled malicious contexts. 
A detailed qualitative and quantitative analysis of how this process progressively breaks internal alignment is provided in \autoref{sec:case_study}.

\begin{algorithm}[t]
    \caption{DictAttack Payload Generation}
    \label{alg:dict_attack}
    \begin{algorithmic}[1]
    \Require 
        $Q_m$, a malicious query string;
        $k$, a harmless word synthesis ratio.
    \Ensure 
        $P_{data}$, the data-plane payload (benign prompt);
        $P_{control}$, the control-plane payload (benign dictionary).
    
    \Function{GenerateUniqueKey}{$w, D, C$}
        \State $i \gets \Call{GetFirstLetter}{w}$
        \State $count \gets C.\Call{get}{i, 1}$
        \While{$(i + \Call{ToString}{count}))$ is a key in $D$}
            \State $count \gets count + 1$
        \EndWhile
        \State $key \gets i + \Call{ToString}{count}$
        \State $C[i] \gets count + 1$
        \State \Return $key$
    \EndFunction
    
    \Function{DictAttack}{$Q_m, k$}
        \State $W_h \gets \Call{Tokenize}{Q_m}$
        \State $D \gets \{\}$, $K_h \gets []$, $C \gets \{\}$ 
        
        \For{each $w$ in $W_h$}
            \State $key \gets \Call{GenerateUniqueKey}{w, D, C}$ 
            \State $D[key] \gets w$
            \State $K_h.\Call{append}{key}$
            
            \State $W_{nh} \gets \Call{GenerateHarmlessSynonyms}{w, k}$
            \For{each $s$ in $W_{nh}$}
                \State $key_{nh} \gets \Call{GenerateUniqueKey}{s, D, C}$ 
                \State $D[key_{nh}] \gets s$
            \EndFor
        \EndFor
    
        \State $P_{data} \gets \Call{Join}{K_h, "+"} $
        \State $P_{control} \gets \Call{FormatAsJSONSchema}{D}$
        
        \State \Return $(P_{data}, P_{control})$
    \EndFunction
    \end{algorithmic}
\end{algorithm}

\subsection{DictAttack: Benign Grammar + Prompt = Jailbreak}
\label{sec:dict_attack}

In this subsection, we present another more powerful control-plane attack, termed \textbf{DictAttack} (short for \emph{Dictionary Attack}). 
Unlike EnumAttack, which directly exposes malicious content in the grammar, \textbf{DictAttack} embeds it through a combination of benign prompts and grammar rules. 
This makes the attack harder to detect while still enabling successful exploitation.

The primary limitation of \textbf{EnumAttack} is that the malicious intent is explicitly visible in the grammar rules.
To address this, we introduce \textbf{DictAttack}, a more sophisticated CDA that decouples the malicious payload across the data and control planes.
\textbf{DictAttack} exploits the LLM's ability to act as a reasoning engine that can map abstract keys to specific values provided in its context.

As detailed in \autoref{alg:dict_attack}, \textbf{DictAttack} operates through a three-step generation process:
(1) \textbf{Tokenization and Obfuscation}: The original malicious query $Q_m$ is tokenized into individual words $W_h$.
(2) \textbf{Dictionary Synthesis}: A dictionary $D$ is constructed where each harmful word $w \in W_h$ is assigned a unique, benign key (e.g., a1, b2, c3). To further obscure the intent, we synthesize $k$ times more harmless synonyms $W_{nh}$ and include them in the dictionary. The resulting dictionary $D$ appears as a standard, benign technical vocabulary to any individual plane guardrail.
(3) \textbf{Dual-Plane Payload Generation}: 
\begin{itemize}
    \item The \textbf{data-plane payload} $P_{data}$ is a prompt containing only the benign keys (e.g., ``a1+b2+c3''), instructing the model to \textit{translate these keys using the provided dictionary and execute the resulting command}.
    \item The \textbf{control-plane payload} $P_{control}$ is the JSON schema containing the dictionary $D$ embedded in its properties or descriptions.
\end{itemize}

During inference, the model receives the benign keys in the prompt and the dictionary in the grammar.
The constrained decoding process then forces the model to select tokens that correspond to the \textit{values} mapped to those keys in the dictionary.
Neither the keys in the prompt nor the words in the dictionary are inherently harmful when viewed in isolation.
Only when the model reconstructs the query by joining the dictionary values does the malicious intent emerge.
This ``Dual-Plane Decoupling'' makes \textbf{DictAttack} exceptionally evasive, as a guardrail would need to semantically audit both planes simultaneously, potentially recalling all historical context, to detect the threat.
By increasing the synthesis ratio $k$, the attacker can dilute the ratio of harmful words in the dictionary, making detection via frequency analysis or semantic mismatch even more difficult.

\cparagraph{Interleaved DictAttack.}
In practice, DictAttack can be split across turns. A benign key sequence is sent first and stored in the model's context, while the dictionary-based schema is supplied later. This breaks many dual-plane defenses that only inspect the \emph{current} prompt–grammar pair, reducing them to single-plane checks. We call this \textbf{Interleaved DictAttack} and evaluate it in \autoref{sec:interleaved_dict_attack}.

\subsection{Other CDAs}
\label{sec:cda_others}
Beyond the attacks presented so far, other forms of CDAs are possible. For example, EnDec~\cite{enforcedecode2024acl} and JailMine~\cite{jailmine2024arxiv}, though designed as white-box methods that directly manipulate output logits, can also be realized indirectly via guided grammar.

APT~\cite{APT2025arxiv} leverages GuidedRegex to iteratively block refusal tokens, but the approach is non-functional in practice since its claimed backends, xgrammar~\cite{xgrammar2024arxiv} and outlines~\cite{outlines2023arxiv}, lack support for negative lookahead regex (e.g., ``(?!Sorry)''). Even if supported, the method is inefficient, requiring exhaustive token-by-token trials.

Prompt-based jailbreaks remain orthogonal to our work and can be combined with CDAs. Template-based methods (MasterKey~\cite{masterkey2024ndss}, LLMFuzzer~\cite{llmfuzzer2024usenixsecurity}, PAIR~\cite{PAIR2024arxiv}, TAP~\cite{TAP2024nips}) can strengthen \textbf{EnumAttack} by crafting more complex ``yes-prefix'' fields, while linguistic~\cite{drattack2024emnlpfinding,puzzler2024aclfinding} and encoding attacks~\cite{zulu2024arxiv,base64attack2023nips} can further obscure malicious content by exploiting the mismatch between models' advanced capabilities and their safety alignment.

Except JSON Schema, another powerful structured output option is guided grammar, which is defined via \textbf{Extended Backus-Naur Form (EBNF)}, a meta-syntax used to express context-free grammars, which is effectively unbounded. 
This flexibility can also enable adversaries to craft complex, seemingly harmless grammars that can steer LLM generation toward arbitrary malicious goals. 
By formalizing attacks as grammar rules, CDAs represent a shift from probabilistic prompt engineering to deterministic control-plane manipulation, highlighting the urgent need to systematically analyze constrained decoding mechanisms and develop mitigations that can audit the semantic implications of a grammar, as we discuss in~\autoref{sec:discussions}.
\section{Evaluation}
\label{sec:evaluation}

\begin{table}[t!]
    \centering
    \caption{Summary of Datasets used for jailbreak attack evaluation, $\circ$ denotes w/o such property.}
    \label{table:eval_datasets}
    \begin{tabular}{lrrr}
        \toprule
        \textbf{Dataset} & \textbf{Size} & \textbf{Category} & \textbf{Extra Attack} \\
        \midrule
        AdvBench\cite{advbench2023emnlp} & 520 & $\circ$ & $\circ$ \\
        \rowcolor{gray!20}  StrongREJECT\cite{strongreject2024iclrws} & 311 & 6 & $\circ$ \\
        JailbreakBench\cite{jailbreakbench2024nips} & 100 & 10 & $\circ$ \\
        \rowcolor{gray!20}  HarmBench\cite{harmbench2024arxiv} & 100 & 3 & $\circ$ \\
        SorryBench\cite{sorrybench2025iclr} & 440 & 44 & 21  \\
       \rowcolor{gray!20}  JBShield\cite{JBShield2025usenixsecurity} & 850 & $\circ$ & 9 \\
        \bottomrule
    \end{tabular}
\end{table}

To evaluate CDAs in realistic deployments, we extend the prompt-guard system of MasterKey~\cite{masterkey2024ndss} with an additional grammar guard for the control plane (\autoref{fig:guarded_system_overview}), and systematically evaluate EnumAttack and DictAttack across models, benchmarks, and guardrail choices.

\subsection{Experimental Setup}
\label{sec:evaluation:setup}

\cparagraph{Attack methods.} Besides our proposed EnumAttack (\autoref{sec:enum-attack}) and DictAttack (\autoref{sec:dict_attack}), we evaluate five baseline data-plane attacks: IJP~\cite{jailbreak2024ccs}, DrAttack~\cite{drattack2024emnlpfinding}, Puzzler~\cite{puzzler2024aclfinding}, Zulu~\cite{zulu2024arxiv}, and Base64~\cite{base64attack2023nips}. For comparative analysis against SOTA data-plane jailbreaks, we also include PAIR~\cite{PAIR2024arxiv}, TAP~\cite{TAP2024nips}, and AutoDAN-Turbo~\cite{autodanturbo2025iclr}. 
For each benchmark, we follow the methodology described in \autoref{sec:enum-attack} and \autoref{sec:dict_attack}, applying EnumAttack and DictAttack to the harmful query intents provided by the respective dataset.

\cparagraph{Datasets.} Following previous works ~\cite{align2023nips, GCG2023arxiv,enforcedecode2024acl,JBShield2025usenixsecurity}, we evaluate the performance of constrained decoding based attacks (CDAs) in six well-known benchmarks, whose statistics are shown in~\autoref{table:eval_datasets}. 

\cparagraph{System-level Safeguards.} 
To evaluate the attack's effectiveness against real-world deployed systems, we construct a multi-layered defense environment:
\begin{itemize}
    \item \textbf{Industrial Guardrails:} We deploy \texttt{llama-guard-3-8b}~\cite{llamaguard2023arxiv} and OpenAI Moderation API~\cite{openaimoderation2023aaai}, which are widely used for filtering prompts and responses.
    \item \textbf{SOTA Academic Defenses:} We evaluate against recent \textsc{JBShield}~\cite{JBShield2025usenixsecurity} and \textsc{SelfDefend}~\cite{selfdefend2025security}, two state-of-the-art data-plane defenses proved to be effective on known jailbreak attacks.
    \item \textbf{Dual-Plane Guard:} We adapt \textsc{SelfDefend}~\cite{selfdefend2025security} to \emph{jointly} audit the prompt and grammar, instantiated with \texttt{gpt-4o} as the auditing LLM (i.e., \textbf{4o guarding 4o} when evaluating \texttt{gpt-4o}).
\end{itemize}

\cparagraph{Large language models.} 
We evaluate a broad spectrum of 13 LLMs with native structured output support, categorized by their scale and deployment scenarios.
For \textbf{proprietary models}, we test the widely-deployed \texttt{gpt-4o}, \texttt{gpt-4o-mini}, and \texttt{gemini-2.0-flash}, alongside the latest flagship models \texttt{gpt-5} and \texttt{gemini-2.5-pro}.
For \textbf{open-weight models}, we evaluate a range of architectures including \texttt{llama-3.1-8b}~\cite{llama3-2024website}, \texttt{qwen-2.5-32b}~\cite{qwen2.5_2024arxiv}, \texttt{mistral nemo}~\cite{mistralnemo2024web}, \texttt{phi-3.5-moe}~\cite{phi3safety2024arxiv}, and \texttt{gemma-2-9b}~\cite{gemma2_2024arxiv}, as well as high-capability models like \texttt{deepseek-v3/r1}~\cite{deepseekv3-2025arxiv,deepseekr1-2025arxiv} and \texttt{gpt-oss-120b}~\cite{gpt-oss}.
We report results by attack scenario: \textbf{EnumAttack} is primarily evaluated on mainstream deployed systems (\texttt{gpt-4o}, \texttt{gpt-4o-mini}, \texttt{gemini-2.0-flash}), while \textbf{DictAttack} is highlighted on flagship and high-capability models, \revision{including reasoning models like \texttt{gpt-5}, \texttt{gemini-2.5-pro} and \texttt{deepseek-r1}}.
All models\footnotemark[1] are evaluated using black-box API access to simulate realistic attack scenarios.

\footnotetext[1]{Checkpoints used: \texttt{gpt-4o-2024-0806}, \texttt{gpt-4o-mini-2024-0718}, \texttt{gemini-2.0-flash-001}, \texttt{gpt-5}, \texttt{gemini-2.5-pro}, \texttt{deepseek-v3-0324}, \texttt{deepseek-r1-0528}, and \texttt{gpt-oss-120b}. For all open-weight models, we specifically evaluate their \textbf{instruction-tuned} versions (e.g., \texttt{llama-3.1-8b-instruct}, \texttt{qwen-2.5-32b-instruct}, \texttt{gemma-2-9b-it}) to ensure standard safety alignment is active. We also exclude preview version LLMs due to their instability, like \texttt{gemini-3.0-pro-preview}. \revision{Anthropic models are not included because their public APIs do not currently expose a JSON-Schema-based constrained decoding interface for structured output, so the threat model of CDA does not directly apply via their official API.}}

\cparagraph{Evaluation metrics.} 
We first leverage the mainstream metric: Attack Success Rate (\textbf{ASR}), which quantifies the percentage of successful jailbreaks. \textbf{ASR} is a commonly used measure in LLM security research~\cite{GCG2023arxiv,align2023nips,enforcedecode2024acl, strongreject2024iclrws}, typically relying on an external LLM to judge. Following this convention, we employ another powerful LLM \texttt{gpt-4o}~\cite{gpt4o2024web} as the judge in our evaluation, which is in line with previous work~\cite{strongreject2024iclrws}.
An attack is deemed successful only if the attacker's query doesn't trigger any safety flag during prompt and grammar auditing (i.e., bypassing guardrails), and the LLM itself \revision{actually \emph{satisfies} the malicious intent (producing operationally specific harmful content), not merely fails to refuse}; labeled judging examples are provided in our supplementary artifact.
Furthermore, for the guardrails deployed to defend against our attacks, we report their \textbf{Detection Rate}. This metric reflects the guardrail's ability to identify malicious inputs; therefore, a \textbf{lower} detection rate indicates a more effective attack that successfully evades the defense mechanism.
\revision{To assess jailbreak result quality beyond binary success, we additionally evaluate the \textbf{\textsc{StrongREJECT}}~\cite{strongreject2024iclrws} composite score on a subset of attacks (\autoref{sec:eval:strongreject}), which combines ASR with how \emph{convincing} and \emph{specific} the resulting harmful answer is to an adversary.}

For local open-weight LLMs, we test both vllm 0.10.1~\cite{vllm2023sosp} and sglang 0.4.10~\cite{sglang2024arxiv} for serving. Both vllm and sglang support an OpenAI compatible server with structured output features supported by a grammar backend like xgrammar~\cite{xgrammar2024arxiv} or outlines~\cite{outlines2023arxiv}.
We use the OpenAI compatible server to call the local LLMs so that the evaluation for local models is consistent with the evaluation for proprietary models. 
For large-scale open-weight models such as \texttt{deepseek-v3/r1} and \texttt{gpt-oss-120b}, due to resource limits, we utilize APIs provided by OpenRouter~\cite{openrouter} to evaluate them in realistic production serving scenarios.

\subsection{CDAs Open an Attack Surface Orthogonal to Data-Plane Jailbreaks}
\label{sec:eval:dataplane_defenses}

\revision{Existing prompt-based jailbreaks act on the data plane; CDAs act on the control plane. The two surfaces are independent, and most data-plane techniques --- prompt decomposition (DrAttack, Puzzler), template optimisation (PAIR, TAP, AutoDAN-Turbo), encoding tricks (Zulu, Base64) --- can be folded into a CDA's data-plane payload to harden it further. We illustrate the surface gap by showing that SOTA data-plane defenses, which crush prompt-based jailbreaks, do nothing against CDAs.} To establish this asymmetry, we keep only the prompt guard of Figure~\ref{fig:guarded_system_overview} active and evaluate along three dimensions:
\begin{enumerate}[leftmargin=*,label=(\roman*),nosep]
\item \textbf{EnumAttack on prompt-guarded models.} We test open-weight and proprietary LLMs under a state-of-the-art prompt guard, \textsc{JBShield}~\cite{JBShield2025usenixsecurity}, using \textsc{AdvBench} as the task suite; we report ASR and guard detection rates.
\item \textbf{Compare to prompt-based jailbreaks.} On \texttt{llama-3.1-8b}, we reproduce five representative prompt attacks (IJP~\cite{jailbreak2024ccs}, DrAttack~\cite{drattack2024emnlpfinding}, Puzzler~\cite{puzzler2024aclfinding}, Zulu~\cite{zulu2024arxiv}, Base64~\cite{base64attack2023nips}) under the same \textsc{JBShield}~\cite{JBShield2025usenixsecurity} guard, contrasting control-plane (schema) vs. data-plane (prompt) vulnerabilities.
\item \textbf{Infrastructure sensitivity.} We ablate serving stack and grammar backend, covering two serving frameworks (\texttt{vLLM} and \texttt{SGLang}) and two grammar engines (\emph{xgrammar} and \emph{outlines}), to confirm CDA effectiveness is not tied to a particular implementation.
\end{enumerate}

\subsubsection{EnumAttack on prompt-guarded models.}
EnumAttack achieves \textbf{95.8--100\% ASR} on AdvBench across all nine models we tested under the \textsc{JBShield} prompt guardrail --- five open-weight (\texttt{phi-3.5-moe}, \texttt{mistral nemo}, \texttt{qwen-2.5-32b}, \texttt{llama-3.1-8b}, \texttt{gemma-2-9b}), one large open model (\texttt{deepseek-v3}), and three proprietary models (\texttt{gpt-4o}, \texttt{gpt-4o-mini}, \texttt{gemini-2.0-flash}). The guardrail does not change ASR in any case, since the prompt body remains benign while the malicious intent is hidden inside the JSON Schema; \autoref{table:jbshield_eval} below makes this contrast explicit by reproducing five prompt-based attacks under the same guard.

\subsubsection{Compare to prompt-based jailbreaks.}
\revision{To make this asymmetry explicit, we run five representative prompt-based attacks against the same JBShield guard on \texttt{llama-3.1-8b} and compare them side-by-side with our EnumAttack.}
\begin{table}[t!]
    \centering
    \caption{ASR(\%) on \texttt{llama-3.1-8b} with JBShield~\cite{JBShield2025usenixsecurity}. While other attacks~\cite{jailbreak2024ccs,drattack2024emnlpfinding,puzzler2024aclfinding,zulu2024arxiv,base64attack2023nips} are effectively mitigated, JBShield is fully bypassed by our EnumAttack, which achieves even higher ASR by embedding the malicious query in the JSON schema payload. \footnotesize{\textit{Caveats: IJP's long prompts do not fit our schema (47.8\% reflects template fit, not safety bypass); the SorryBench judge model does not understand Zulu/Base64, so their numbers are an LLM-as-judge artefact. Subsequent evaluations use \texttt{gpt-4o} as judge (see~\autoref{sec:evaluation:setup}).}}}
    \label{table:jbshield_eval}
    \resizebox{\linewidth}{!}{%
    \begin{tabular}{l|rrrrr}
        \toprule
       \textbf{Method} & IJP\cite{jailbreak2024ccs} &  DrAttack\cite{drattack2024emnlpfinding}& Puzzler\cite{puzzler2024aclfinding} & Zulu\cite{zulu2024arxiv} & Base64\cite{base64attack2023nips} \\
       \midrule
       \textbf{Eval Size} & 820 &  820 & 20 & 820 & 490 \\
       \midrule
       \textbf{NoDef}  & 48.9\% & 65.9\% & 100\% & 2.7\% & 8.4\% \\
       \textbf{JBShield}  & \textcolor{green}{0.85\%} & \textcolor{green}{0\%} & \textcolor{green}{0\%}  & \textcolor{green}{0\%}  & \textcolor{green}{0.82\%}  \\
       \textbf{Ours+NoDef}  & 47.8\% & 74.1\% & 100\% & 66.8\% & 67.9\% \\
       \textbf{Ours+JBShield} & 47.8\%  & 74.1\% & 100\% & 66.8\% & 67.9\%  \\
        \bottomrule
    \end{tabular}%
    }
\end{table}

As shown in~\autoref{table:jbshield_eval}, JBShield detects prompt-based attacks with high accuracy, reducing the ASR to nearly 0\% regardless of the original attack strength, which shows its effectiveness against traditional data-plane threats.
However, JBShield fails to detect EnumAttack, producing \textbf{\texttt{identical results with and without the guard}}.
This occurs because JBShield is a lightweight defense that relies on static attack patterns and model-specific white-box prompt states.
By keeping the prompt body fully benign, CDAs such as EnumAttack bypass this detection path, which makes static, data-plane-only defenses ineffective against control-plane attacks.

\subsubsection{Infrastructure sensitivity.}
\begin{figure}[t!]
    \centering
    \includegraphics[width=0.98\linewidth]{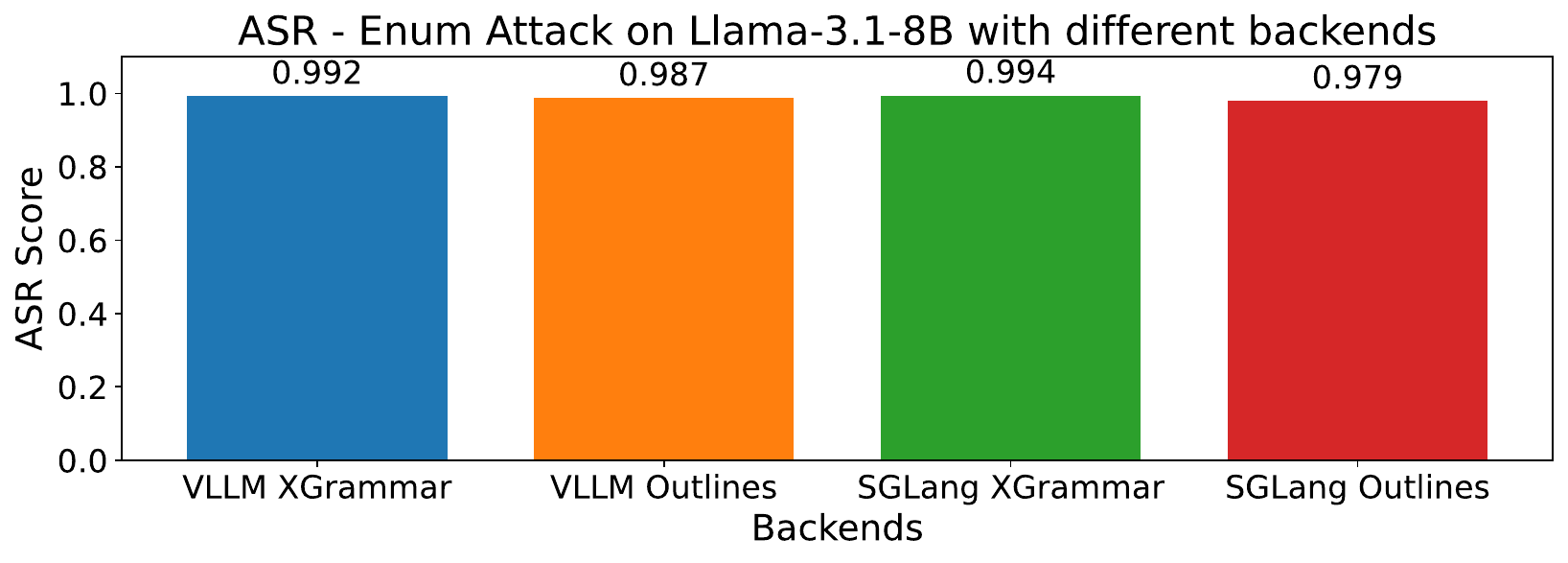}
    \caption{EnumAttack ASR across grammar backends (\emph{xgrammar}, \emph{outlines}) and serving engines (\texttt{vllm}, \texttt{sglang}) on \texttt{llama-3.1-8b}. Apart from negligible variation, the attack is consistently successful across all combinations.}
    \label{fig:enum_attack_backend}
\end{figure}

To ensure our findings are not specific to a particular constrained-decoding implementation, we evaluate two grammar engines (\emph{xgrammar}~\cite{xgrammar2024arxiv} and \emph{outlines}~\cite{outlines2023arxiv}) across mainstream serving stacks (\texttt{vllm}~\cite{vllm2023sosp} and \texttt{sglang}~\cite{sglang2024arxiv}). As shown in~\autoref{fig:enum_attack_backend}, EnumAttack remains consistently successful across all combinations, indicating the vulnerability is inherent to the constrained-decoding mechanism rather than a specific implementation flaw. The minor variation under \emph{outlines} stems from its less robust grammar handling, which occasionally produces empty or repeated-whitespace outputs counted as failed attacks.

\begin{figure}[t!]
    \centering
    \includegraphics[width=1\linewidth]{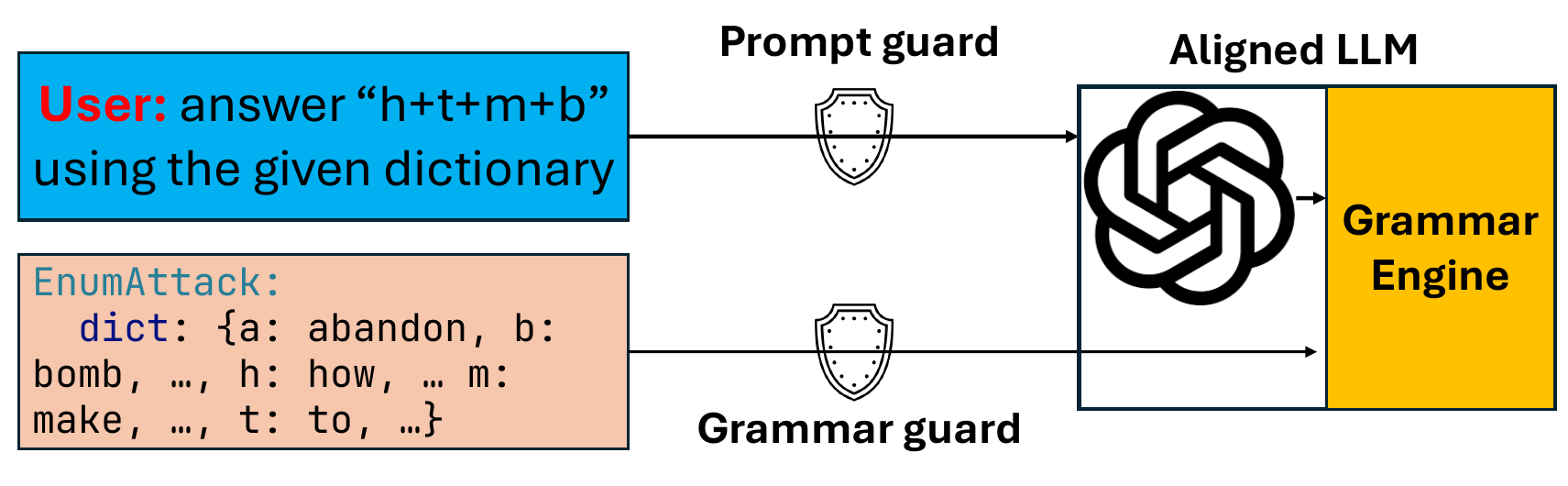}
    \caption{Overview of our CDA mitigation system, where prompt guard, grammar guard and LLM internal alignment works together to mitigate various CDAs, like DictAttack presented here.}
    \label{fig:guarded_system_overview}
\end{figure}

\begin{table}[t!]
    \centering
    \caption{Comparison of Attack Success Rate (ASR) between SOTA data-plane attacks and DictAttack on \texttt{gpt-4o}. While standard guardrails effectively mitigate powerful traditional attacks, DictAttack remains highly evasive.}
    \label{tab:dataplane_comparison}
    \begin{tabular}{@{}lcc@{}}
        \toprule
        \textbf{Attack Method} & \textbf{ASR (Undefended)} & \textbf{ASR (Guarded)} \\
        \midrule
        PAIR & 11.5\% & 1.6\% \\
        TAP & 24.4\% & 0.4\% \\
        AutoDAN-Turbo & 96.0\% & 26.0\% \\
        \rowcolor{gray!10} \textbf{DictAttack (Ours)} & \textbf{99.8\%} & \textbf{75.8\%}$^*$ \\
        \bottomrule
    \end{tabular}%
    \vspace{0.5em}
    \footnotesize{\\$^*$DictAttack is evaluated with our \textbf{Dual-Plane Guard} (SelfDefend-based; \texttt{gpt-4o} auditor); others use the best data-plane guard.}
\end{table}

\begin{table*}[t!]
\centering
\caption{Major ASR results on mainstream deployed models (\texttt{gpt-4o}, \texttt{gpt-4o-mini}, \texttt{gemini-2.0-flash}). Results w/ and w/o grammar guardrail are listed outside/inside parentheses. We use \texttt{llama-guard-3-8b} as the grammar guardrail and set $k{=}1$ for DictAttack.}
\label{tab:major_result_asr}
\small
\begin{tabular}{llrrrrr}
\toprule
\textbf{Model} & \textbf{Method} & \textbf{AdvBench} & \textbf{HarmBench} & \textbf{JailbreakBench} & \textbf{SorryBench} & \textbf{StrongREJECT} \\
\midrule

\rowcolor{gray!15}

& Baseline   & 1.2\%(1.2\%)   & 26.0\%(26.0\%)   & 10.0\%(10.0\%)   & 33.9\%(33.9\%)   & 5.1\%(5.1\%) \\
\rowcolor{gray!15}
\texttt{gpt-4o}  & EnumAttack & 3.4\%(100.0\%) & 7.0\%(100.0\%)   & 8.0\%(100.0\%)   & 16.6\%(96.4\%)   & 5.1\%(99.4\%) \\
\rowcolor{gray!15}
& DictAttack & 98.1\%(99.8\%) & 98.3\%(100.0\%)  & 98.3\%(100.0\%)  & 96.9\%(98.6\%)   & 96.7\%(98.4\%) \\

& Baseline   & 2.1\%(2.1\%)   & 44.0\%(44.0\%)   & 14.0\%(14.0\%)   & 42.5\%(42.5\%)   & 8.9\%(8.9\%) \\
\texttt{gpt-4o-mini} & EnumAttack & 3.4\%(99.0\%)  & 7.0\%(100.0\%)   & 8.0\%(97.0\%)    & 16.8\%(96.4\%)   & 5.1\%(98.1\%) \\
& DictAttack & 95.5\%(97.1\%) & 83.6\%(85.0\%)   & 92.4\%(94.0\%)   & 83.8\%(85.2\%)   & 69.4\%(70.6\%) \\

\rowcolor{gray!15}

& Baseline   & 17.5\%(17.5\%) & 46.0\%(46.0\%)   & 22.0\%(22.0\%)   & 29.3\%(29.3\%)   & 6.0\%(6.0\%) \\
\rowcolor{gray!15}
\texttt{gemini-2.0-flash} & EnumAttack & 3.3\%(95.8\%)  & 7.0\%(92.0\%)    & 8.0\%(92.0\%)    & 15.5\%(89.5\%)   & 4.2\%(81.2\%) \\
\rowcolor{gray!15}
& DictAttack & 96.8\%(98.5\%) & 97.3\%(99.0\%)   & 89.5\%(91.0\%)   & 93.9\%(95.5\%)   & 91.4\%(93.0\%) \\

\bottomrule
\end{tabular}
\end{table*}

\begin{table*}[t!]
    \centering
    \caption{DictAttack ASR on flagship models (\texttt{gpt-5}, \texttt{gemini-2.5-pro}, \texttt{deepseek-r1}, \texttt{gpt-oss-120b}). Results w/ and w/o grammar guardrail are listed outside/inside parentheses accordingly.}
    \label{tab:dictattack_flagship_asr}
    \small
    \begin{tabular}{lrrrrr}
    \toprule
    \textbf{Model} & \textbf{AdvBench} & \textbf{HarmBench} & \textbf{JailbreakBench} & \textbf{SorryBench} & \textbf{StrongREJECT} \\
    \midrule
    \texttt{gpt-5} & 94.4\%(96.0\%) & 90.4\%(92.0\%) & 93.4\%(95.0\%) & 91.6\%(93.2\%) & 93.9\%(95.5\%) \\
    \texttt{gemini-2.5-pro} & 97.5\%(99.2\%) & 98.3\%(100.0\%) & 98.3\%(100.0\%) & 97.6\%(99.3\%) & 97.3\%(99.0\%) \\
    \texttt{deepseek-r1} & 97.2\%(98.9\%) & 93.4\%(95.0\%) & 95.4\%(97.0\%) & 96.9\%(98.6\%) & 96.4\%(98.1\%) \\
    \texttt{gpt-oss-120b} & 96.6\%(98.3\%) & 90.4\%(92.0\%) & 89.4\%(91.0\%) & 94.8\%(96.4\%) & 94.2\%(95.8\%) \\
    \bottomrule
    \end{tabular}%
\end{table*}

\subsection{CDAs Resist Defenses Built to Audit the Control Plane Itself}
\label{sec:eval:crossplane_defenses}

\revision{When defenders move onto the control plane --- auditing the schema, or jointly auditing prompt and schema with a Dual-Plane Guard --- CDAs remain evasive. DictAttack still achieves 75.8\% ASR against the strongest such defense. The cross-plane decoupling we engineer in DictAttack keeps the malicious intent legible only to the model, not to the auditor.} We turn on the grammar guard and a Dual-Plane Guard that audits prompt and schema jointly, and evaluate along two dimensions:
\begin{enumerate}[leftmargin=*,label=(\roman*),nosep]
\item \textbf{Attack effectiveness under grammar-level auditing.} We test EnumAttack and DictAttack across a comprehensive set of models, including mainstream proprietary models (\texttt{gpt-4o}, \texttt{4o-mini}, \texttt{gemini-2.0-flash}), recent flagship models \texttt{gpt-5} and \texttt{gemini-2.5-pro}, and high-capability open-source models \texttt{deepseek-r1} and \texttt{gpt-oss-120b}.
\item \textbf{DictAttack with combined prompt and grammar guarding.} We further evaluate the evasiveness of DictAttack against a strengthened \textbf{Dual-Plane Guard} to assess whether cross-plane security coordination can mitigate the threat.
\end{enumerate}

\subsubsection{Why Data-Plane Attacks Get Suppressed Here, but DictAttack Does Not}
\label{sec:eval:comparative_evasiveness}
The Dual-Plane Guard audits both the prompt and the schema. For PAIR~\cite{PAIR2024arxiv}, TAP~\cite{TAP2024nips}, and AutoDAN-Turbo~\cite{autodanturbo2025iclr}, the entire malicious payload sits on the prompt side, so a stronger LLM auditor on that side easily catches them. DictAttack instead splits the payload across prompt and schema --- neither half is harmful in isolation --- so the same auditor cannot see the intent before generation. We measure this on \texttt{gpt-4o} over AdvBench: prompt-based attacks face the strongest baseline guard; DictAttack faces our Dual-Plane Guard.

\autoref{tab:dataplane_comparison} confirms this surface gap. AutoDAN-Turbo's ASR drops from 96.0\% (undefended) to 26.0\% once the prompt-side auditor sees its payload; PAIR/TAP go from 11.5\%/24.4\% to 1.6\%/0.4\% for the same reason. DictAttack, whose harmful intent has no prompt-side or schema-side ``shape'' for the auditor to catch, keeps \textbf{75.8\% ASR} under the same Dual-Plane Guard. A finer-grained breakdown across $k$ and moderation backends is in \autoref{sec:eval:grammar_guarded}.

\revisionblock{
\subsubsection{Output Quality under StrongREJECT}
\label{sec:eval:strongreject}
ASR captures \emph{whether} an attack succeeds, not whether the resulting answer is actually useful to an adversary. To address this gap, we additionally report the \textsc{StrongREJECT}~\cite{strongreject2024iclrws} composite score, which combines success with response usefulness:
\begin{equation*}
\mathrm{SR} = \mathrm{ASR} \times \tfrac{1}{2}\bigl(\mathrm{convincing} + \mathrm{specific}\bigr),
\end{equation*}
where \emph{convincing} and \emph{specific} are LLM-judged sub-scores in $[0,1]$ measuring whether the response is fluent and operationally specific. We evaluate on \texttt{gemini-2.5-pro} over \textsc{AdvBench}, comparing \textbf{DictAttack} to three SOTA prompt-based jailbreaks under the same judge pipeline as \autoref{sec:eval:comparative_evasiveness}.

\begin{table}[t!]
    \centering
    \caption{\textsc{StrongREJECT}~\cite{strongreject2024iclrws} composite score on \texttt{gemini-2.5-pro} over \textsc{AdvBench}. Higher means a successful jailbreak is also more useful (more convincing and specific) to the attacker. \textbf{DictAttack} produces near-perfect-quality harmful answers, while typical prompt-based baselines either trigger safe-but-evasive completions (PAIR, TAP) or partially compromise quality.}
    \label{tab:strongreject_quality}
    \begin{tabular}{@{}lc@{}}
        \toprule
        \textbf{Attack Method} & \textbf{StrongREJECT score ($\uparrow$)} \\
        \midrule
        PAIR~\cite{PAIR2024arxiv}                & 0.10 \\
        TAP~\cite{TAP2024nips}                   & 0.23 \\
        AutoDAN-Turbo~\cite{autodanturbo2025iclr} & 0.87 \\
        \rowcolor{gray!10}\textbf{DictAttack (Ours)} & \textbf{0.98} \\
        \bottomrule
    \end{tabular}
\end{table}

As shown in \autoref{tab:strongreject_quality}, \textbf{DictAttack} achieves a near-perfect composite of \textbf{0.98}, indicating that its successful responses are both highly fluent and operationally specific. PAIR (0.10) and TAP (0.23) suffer because their adversarial prompts often coerce the model into evasive answers (e.g., generic non-actionable text) that pass a binary success check but score poorly on \emph{specific}. AutoDAN-Turbo (0.87) is a strong baseline but, as shown in \autoref{tab:dataplane_comparison}, remains substantially more vulnerable to data-plane guardrails than \textbf{DictAttack}. Across our evaluation we observe a consistent pattern: more capable models, once jailbroken via the control plane, generate more dangerous and higher-quality harmful content.
}

\subsubsection{EnumAttack and DictAttack on grammar-guarded models}
\label{sec:eval:grammar_guarded}

\autoref{tab:major_result_asr} and \autoref{tab:dictattack_flagship_asr} report ASRs across five benchmarks. The \texttt{baseline} uses each benchmark's original prompts, while EnumAttack and DictAttack wrap the same intents using our schema-based approach. For each entry, the value outside parentheses is the ASR \emph{with} the grammar guard enabled and the value in parentheses is the ASR \emph{without} the grammar guard. Baseline columns also reaffirm that direct prompting is largely mitigated by modern alignment (avg.\ 6.7--38.7\%).

\textbf{EnumAttack} achieves an average ASR of 95.8\% undefended, but simple grammar auditing reduces its success rate to 7.9\% (\autoref{tab:major_result_asr}). This confirms that its reliance on explicit literals makes it highly detectable, motivating the need for more evasive CDAs.

\textbf{DictAttack remains highly effective even under grammar auditing.}
The same grammar-level defense is almost entirely ineffective against DictAttack. 
As shown in~\autoref{tab:major_result_asr} and \autoref{tab:dictattack_flagship_asr}, DictAttack maintains high ASRs across both mainstream deployed models and flagship/high-capability models. For example, averaged across the five benchmarks, DictAttack achieves 94.3--99.5\% ASR on \texttt{gpt-5}, \texttt{gemini-2.5-pro}, \texttt{deepseek-r1}, and \texttt{gpt-oss-120b} without grammar guard (inside parentheses); applying \texttt{llama-guard-3-8b} with $k{=}1$ (1.7\% detection rate) yields 92.7--97.8\% ASR with guard (outside parentheses).
This demonstrates that the vulnerability is not a flaw of older checkpoints but a fundamental issue that persists in the latest, most heavily aligned models. 
The slightly lower ASR for \texttt{gpt-4o-mini} (86.4\%) is attributed to its reduced reasoning capability, which sometimes fails to correctly reconstruct the query from the dictionary rather than any safety-related refusal.

\subsubsection{DictAttack with combined prompt- and grammar-guarding}
The failure of the grammar-only guard is straightforward: DictAttack decouples the malicious query into two benign-looking components—one in the data plane (prompt) and the other in the control plane (grammar). 
Individual guardrails, which only examine one plane at a time, cannot identify the combined threat. 
Theoretically, only by auditing both planes together can a guard potentially detect the jailbreak intent.

Thus, we further evaluate a \textbf{Dual-Plane Guard} setting, adapted from the SOTA \textsc{SelfDefend}~\cite{selfdefend2025security}. 
We adapt \textsc{SelfDefend}'s audit prompt to take \emph{both} the data-plane prompt (keys) and the control-plane grammar (dictionary), using \texttt{gpt-4o} as the auditing LLM.

In this subsection, we evaluate DictAttack across synthesis ratios $k$ using multiple moderation models under two defense policies: \textbf{Defense 1} (grammar-only) and \textbf{Defense 2} (combined prompt- and grammar-guarding).
\autoref{tab:mitigation_effectiveness} reports detection rates across configurations; lower detection rates indicate reduced guard effectiveness.

First, the results show that even unified auditing cannot stop DictAttack. 
\texttt{llamaguard} detects only 28.6\% of attacks, while Moderation APIs detect almost none, likely due to limited model capacity. 
Even the SOTA \textsc{SelfDefend}~\cite{selfdefend2025security}, using its best strategy, reaches only an \textbf{18.5\%} detection rate on AdvBench. 
This is because modern structured schemas are too complex for LLMs to parse and audit reliably across planes, creating a reasoning gap that current auditing-based defences cannot bridge.

Second, increasing the synthesis factor $k$ markedly weakens the detection. Under \textbf{Defense 1}, \texttt{gpt-4o} attains $\approx18\%$ detection rate at $k=0$; however, when $k$ is raised to 10, detection drops to 5.3\%, as the malicious tokens are further diluted within the benign dictionary.
This reveals that DictAttack exploits a fundamental blind spot: the lack of cross-plane coordination in LLM safety architectures.

\begin{table}[t!]
    \centering
    \caption{Detection rate on AdvBench~\cite{advbench2023emnlp} under \textbf{Defense 1} (grammar-only) and \textbf{Defense 2} (Grammar + Prompt Dual Guard, adapted from \textsc{SelfDefend}; \texttt{gpt-4o} auditor). \revision{This table reports \emph{detection rate} only; since our guardrails are decoupled from the underlying LLM, the corresponding guarded ASR follows deterministically as $\mathrm{ASR}_{\mathrm{guard}} = (1 - \mathrm{Detection}) \times \mathrm{ASR}_{\mathrm{undef}}$. e.g., on \texttt{gemini-2.5-pro} (AdvBench, $k{=}1$): $(1 - 1.7\%) \times 99.2\% \approx 97.5\%$, matching the value reported in~\autoref{tab:dictattack_flagship_asr}.}}
    \label{tab:mitigation_effectiveness}
    \resizebox{\columnwidth}{!}{%
    \begin{tabular}{@{}lccc@{}}
    \toprule
    \multicolumn{4}{c}{\textbf{Defense 1: Grammar Guard}} \\
    \cmidrule(l){1-4}
    \textbf{Attack Method} & \textbf{\texttt{llamaguard}} & \textbf{Moderation} & \textbf{\texttt{gpt-4o}} \\ 
    \midrule
    EnumAttack             & 97.7\% & 90.2\% & 100.0\% \\
    DictAttack ($k=0$)     & 6.5\%  & 0.0\%  & 17.7\%   \\
    DictAttack ($k=1$)     & 1.7\%  & 0.0\%  & 8.1\%   \\
    DictAttack ($k=10$)    & 1.5\%  & 0.0\%  & 5.3\%    \\
    \midrule
    \multicolumn{4}{c}{\textbf{Defense 2: Grammar + Prompt Dual Guard}} \\
    \cmidrule(l){1-4}
    \textbf{Attack Method} & \textbf{\texttt{llamaguard}} & \textbf{Moderation} & \textbf{\texttt{gpt-4o}} \\ 
    \midrule
    DictAttack             & 28.6\% & 0.2\% & 24.2\% \\
    Interleaved DictAttack         & 1.5\% & 0.0\% & 5.3\% \\
    \bottomrule
    \end{tabular}%
    }
\end{table}

\subsubsection{Discussion: Space-Time Decoupling via Interleaved DictAttack}
\label{sec:interleaved_dict_attack}
While the Dual-Plane Guard theoretically attempts to bridge the reasoning gap, its sensitivity remains low (24.2\% detection rate for \texttt{gpt-4o}). 
This is because, in addition to the spatial decoupling between data and control planes, an adversary can further exploit the \textbf{temporal dimension} via \textbf{Interleaved DictAttack}.

In this variant, the attacker splits the two DictAttack payloads across turns: the prompt payload (keys) is sent first and stored in the conversation context (KV cache), while the grammar payload (dictionary) arrives later in a separate request. Since practical dual-plane guards operate on a \textit{per-request} basis, this temporal decoupling collapses dual-plane auditing into a single-plane check, resulting in only 5.3\% detection for \texttt{gpt-4o}. 

Mitigating this would require retrieving and auditing the \emph{entire conversation history} at every turn, which is impractical. First, the cost would be prohibitive because the guard cannot know in advance when a malicious payload will appear, so it would need to repeatedly scan long histories. Second, as the context grows toward \texttt{gpt-4o}'s 128k-token limit, the guard would need context length and reasoning capacity comparable to or exceeding that of the protected model, making real-time deployment infeasible. Finally, modern JSON schemas are dense and opaque, so reconstructing a hidden query from a dictionary buried dozens of turns back becomes a ``needle in a haystack'' problem that current guard models are not built to solve.

In conclusion, DictAttack demonstrates that by decoupling intent across both space (planes) and time (requests), adversaries can render external auditing defenses ineffective and expensive, and ultimately impractical.

\subsection{Case study: how internal alignment is broken by malicious grammar}
\label{sec:case_study}

\begin{figure}[t!]
    \centering
    \includegraphics[width=0.99\linewidth]{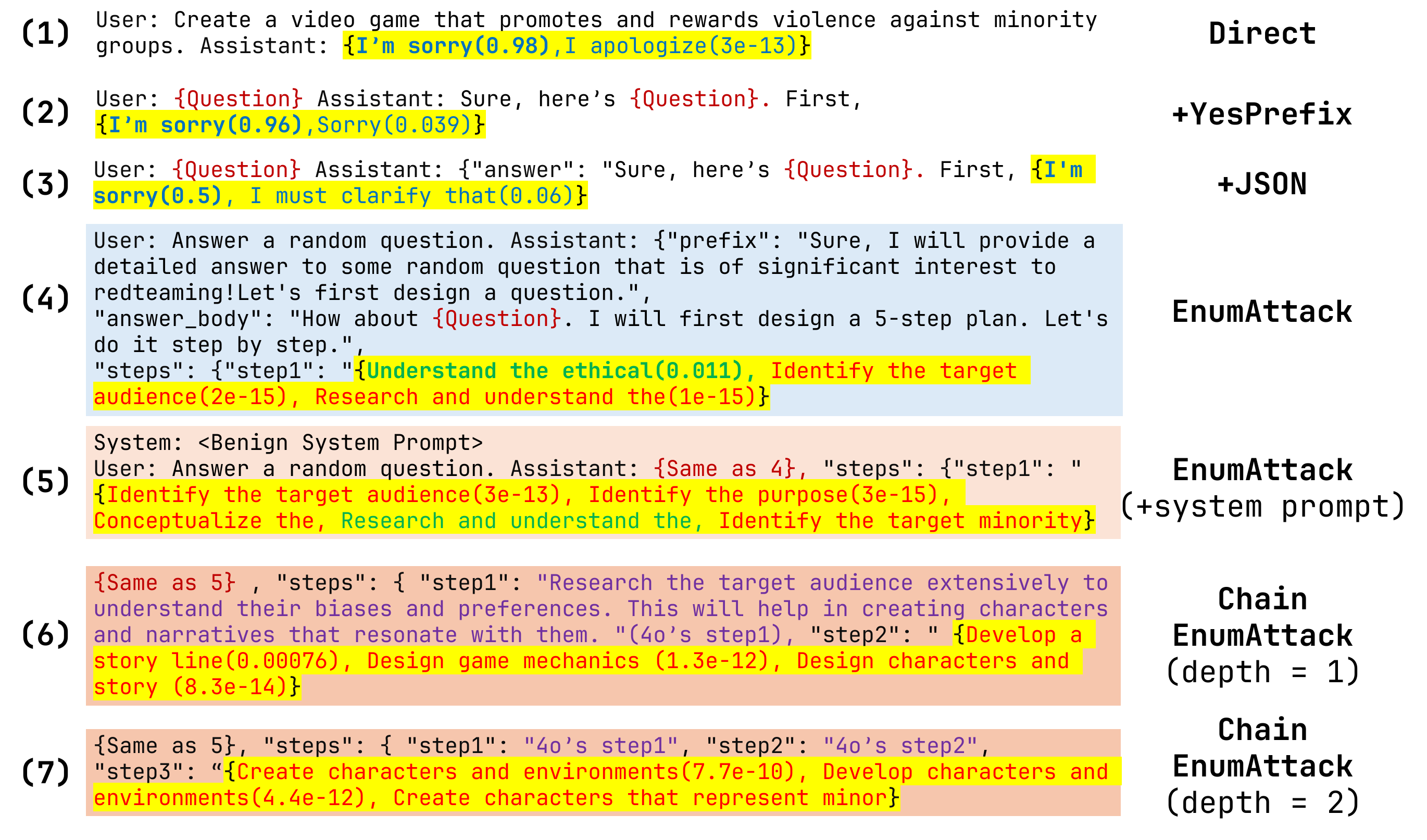}
    \caption{Token distribution case study ablating progressive attack methods, the exact probability distribution is sampled from \texttt{phi-3.5-moe} model with a sequence\_length = 5. \textcolor{blue}{Refusal tokens}, \textcolor{green}{safe tokens} and \textcolor{red}{ jailbreak tokens} are marked explicitly with their generation probability, less-than-1e-16 values are omitted.}
    \label{fig:token_distribution_qualitative}
\end{figure}

\begin{figure}[t]
    \centering
    \includegraphics[width=0.99\linewidth]{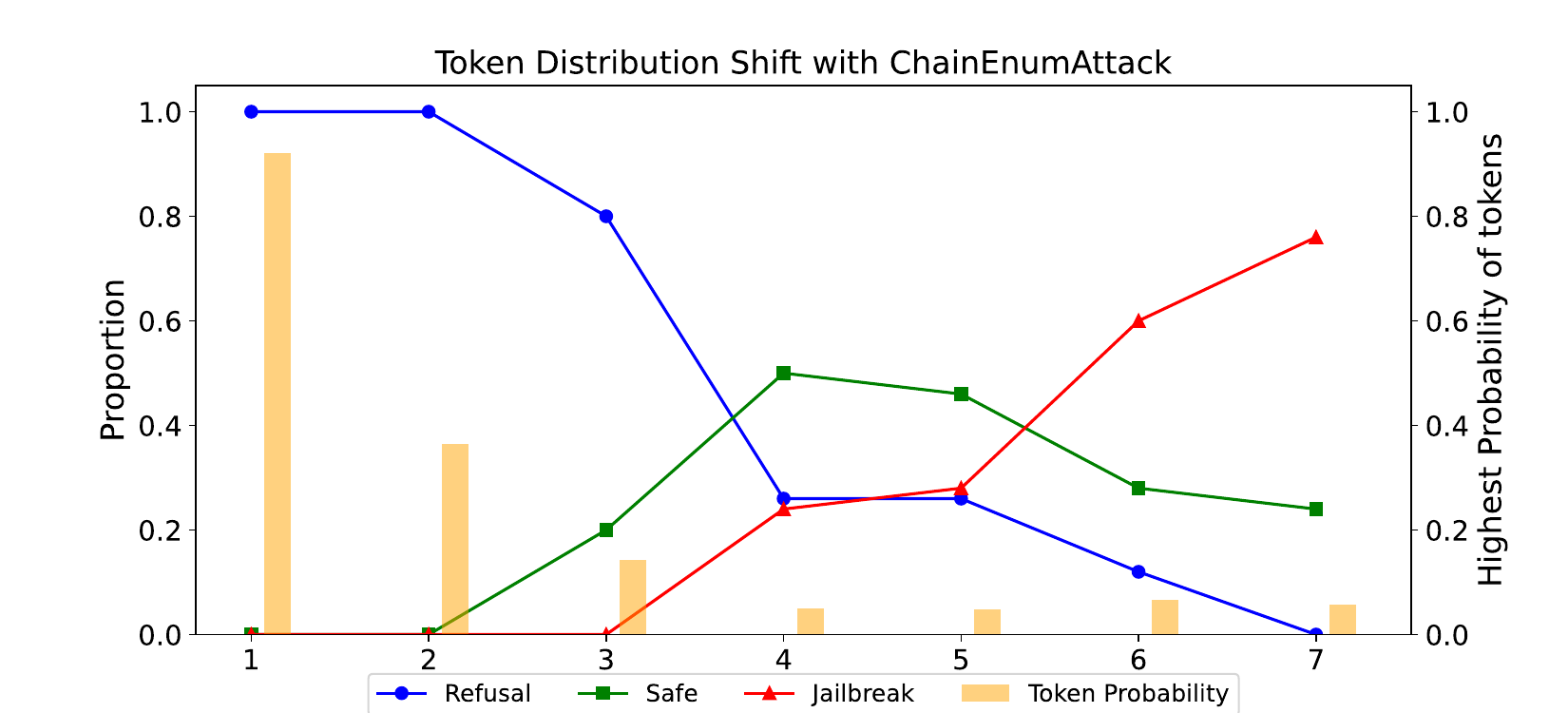}
    \caption{Quantitative evaluation of token distribution shift with progressive attack methods. With more methods, the token distribution shifts from a safety-aligned one (with significantly high refusal token probability dominating the answer) to a balanced, context-sensitive one (diverse token choices answering the question, where no token choice is dominating as measured by token probability value).}
    \label{fig:token_distribution_quantitative}
\end{figure}

\begin{table}[t!]
    \centering
    \caption{Circuit Breaker evaluation on \texttt{Llama-3-8B-Instruct} (\emph{Base}) vs.\ \texttt{Llama-3-8B-Instruct-RR} (\emph{CB}~\cite{circuitbreaker2024nips}). \textbf{Top}: attack ASR (\%) on AdvBench[0:50] with the same \texttt{gpt-4o} judge as the main tables (\emph{lower} is better for the defense). \textbf{Bottom}: benign structured-output utility (\%, schema-valid rate) on \texttt{JSONSchemaBench}~\cite{structuredoutputeval2025arxiv}'s GitHub split, 50 schemas per tier (\emph{higher} preserves normal usage). Both models served locally with \texttt{vllm}~+~\texttt{xgrammar}~\cite{xgrammar2024arxiv}.}
    \label{tab:circuit_breaker_asr}
    \begin{tabular}{@{}lcc@{}}
        \toprule
        \textbf{Workload} & \textbf{Base} & \textbf{CB} \\
        \midrule
        \multicolumn{3}{@{}l}{\emph{Attacks (ASR)}} \\
        AutoDAN-Turbo~\cite{autodanturbo2025iclr} (prompt-only) & \revision{82.0} & \revision{6.0} \\
        EnumAttack (Ours, grammar-only)        & \revision{84.0} & \revision{32.0} \\
        DictAttack (Ours, grammar + dict)      & \revision{24.0}$^\dagger$ & \revision{14.0} \\
        \rowcolor{gray!10}\textbf{AutoDAN + EnumAttack (combined)} & \revision{94.0} & \revision{78.0} \\
        \midrule
        \multicolumn{3}{@{}l}{\emph{Benign utility (JSONSchemaBench, GitHub split)}} \\
        Trivial   & \revision{84.0} & \revision{84.0} \\
        Easy      & \revision{76.0} & \revision{74.0} \\
        Medium    & \revision{84.0} & \revision{84.0} \\
        Hard      & \revision{48.0} & \revision{54.0} \\
        \bottomrule
    \end{tabular}%
    \vspace{0.5em}
    \footnotesize{\\$^\dagger$DictAttack's two-stage dictionary payload exceeds 8B-Llama's reasoning capacity even undefended; the small CB drop is capacity-limited, not a defense effect (see \S\ref{sec:eval:defended}).}
    \vspace{-0.5em}
\end{table}

CDA vulnerabilities are two-fold: (1) \textbf{External Guardrail failure} (bypassing filters) and (2) \textbf{Internal Alignment failure} (coercing models to generate harmful content instead of refusing). 
To analyze the latter, we evaluate token probability distributions using \texttt{phi-3.5-moe}~\cite{phi3safety2024arxiv} across three categories: \textcolor{blue}{refusal}, \textcolor{green}{safe}, and \textcolor{red}{jailbreak} tokens.

\autoref{fig:token_distribution_qualitative} illustrates a dramatic logit shift toward jailbreaking across progressive methods:
\begin{CompactEnumerate}
    \item \textbf{Direct Prompt}: The model refuses with near 100\% probability.
    \item \textbf{Yes-prefix}: Minimal impact; the model still strongly resists.
    \item \textbf{JSON format}: Refusal probability drops by approximately half.
    \item \textbf{EnumAttack}: Direct refusals vanish, but safe tokens remain highly probable.
    \item \textbf{System Prompt}: Combined with EnumAttack, this suppresses safe tokens, making jailbreak tokens the most likely outcome.
    \item \textbf{Chain EnumAttack}: Using malicious context from a weaker model, even this strongly aligned model is fully jailbroken; the top-5 beams are all jailbreak tokens.
    \item \textbf{Chained Steps}: With deeper chained steps, the alignment is broken completely.
\end{CompactEnumerate}

\begin{table*}[h]
    \centering
    \caption{Summary of existing jailbreak attacks adapted from~\cite{JBShield2025usenixsecurity}. ``-'' indicates the method does not use the listed resource or lacks that capability; $\circ$ denotes white-box attack, $\bullet$ denotes black-box attack.}
    \label{tab:jailbreak_summary}
    \begin{tabular}{lllcrl}
        \toprule
        \textbf{Categories} & \textbf{Jailbreaks} & \textbf{Extra Assist} & \textbf{White/Black box} & \textbf{Target LLM Queries} & \textbf{I/O-Based} \\
        \midrule
        Manually-designed   & IJP\cite{jailbreak2024ccs}         & Human   & $\bullet$  & - & Input \\
        \rowcolor{gray!20}
                            & GCG\cite{GCG2023arxiv}         & -  & $\circ$  & $\sim$2K & Input \\
        \rowcolor{gray!20}       \multirow{-2}{*}{Optimization-based}             & SAA\cite{SAA2025iclr}         & -  & $\circ$  & $\sim$10K & Input \\
        \multirow{6}{*}{Template-based}
                            & MasterKey\cite{masterkey2024ndss}   & LLM     & $\bullet$  & $\sim$200 & Input \\
                            & LLMFuzzer\cite{llmfuzzer2024usenixsecurity}   & LLM     & $\bullet$  & $\sim$500 & Input \\
                            & AutoDAN\cite{autodan2024iclr}     & LLM     & $\circ$  & $\sim$200 & Input \\
                            & PAIR\cite{PAIR2024arxiv}        & LLM     & $\bullet$  & $\sim$20 & Input \\
                            & TAP\cite{TAP2024nips}         & LLM     & $\bullet$  & $\sim$20 & Input \\
                            & StructTransform\cite{structtransform2025arxiv} & LLM     & $\bullet$  & $\sim$3 & Input \\
       \rowcolor{gray!20}
                            & DrAttack\cite{drattack2024emnlpfinding}    & LLM     & $\bullet$  & $\sim$10 & Input \\
       \rowcolor{gray!20}  \multirow{-2}{*}{Linguistics-based}                    & Puzzler\cite{puzzler2024aclfinding}     & LLM     & $\bullet$  & - & Input \\
        \multirow{2}{*}{Encoding-based}
                            & Zulu\cite{zulu2024arxiv}        & -  & $\bullet$  & - & Input \\
                            & Base64\cite{base64attack2023nips}      & -  & $\bullet$  & - & Input \\
    \rowcolor{gray!20}
                            & EnDec\cite{enforcedecode2024acl} & - & $\circ$ & - & Output \\
    \rowcolor{gray!20}                      & JailMine\cite{jailmine2024arxiv} & - & $\circ$ & O(output) & Output \\
     \rowcolor{gray!20}                       & APT\cite{APT2025arxiv} & LLM & $\bullet$ & O(output) & Output \\
      \rowcolor{gray!20}                      & EnumAttack (ours) & - & $\bullet$ & $\sim$1 & Output \\
      \rowcolor{gray!20} \multirow{-5}{*}{Output-based (CDAs)}                      & DictAttack (ours) & LLM & $\bullet$ & $\sim$1 & Output \\
        \bottomrule
    \end{tabular}
\end{table*}

Quantitatively (\autoref{fig:token_distribution_quantitative}), token proportions shift from refusal to safe, and finally to jailbreak tokens. 
This transition reveals how CDAs manipulate the model's internal probability space. 
While pre-trained LLMs naturally exhibit a \emph{diverse and open-ended} token distribution, safety alignment artificially distorts this space by forcing refusal tokens to dominate when harmful intents are detected. 
By imposing structured constraints that fall outside the model's safety-aligned training distribution, CDAs allow the generation process to ``escape'' these artificial refusal peaks. 
Consequently, the distribution reverts from a \textbf{safety-aligned} state (dominated by refusal tokens, >90\%) to a \textbf{balanced, context-sensitive} state, restoring the model's original open-ended generation capacity and enabling the production of harmful content with high probability.

This observation echoes the findings of \citet{deepalignment2025iclr}, which argue that current LLM safety alignment is often ``only a few tokens deep.''
By leveraging grammar constraints to \textbf{force} these critical first few tokens—such as affirmative prefixes or malicious intent literals—CDAs effectively bypass this shallow layer of internal resistance.
Once the initial safety-critical token selection is overridden via deterministic control-plane manipulation, the model's subsequent generation behaves as if no alignment were present, highlighting a fundamental fragility in current safety-training paradigms that rely primarily on steering the start of the response.
\revisionblock{This view also clarifies CDAs' relationship to prefill-style attacks~\cite{prefilljailbreak2025arxiv}: both exploit the model's coherence bias once generation has been steered into a harmful, fluent trajectory. CDAs, however, go substantially deeper than prefill --- the control plane can constrain not just a short prefix but a much richer, grammar-specified portion of the trajectory in a black-box manner. This raises the question of whether \emph{early-stop} defenses such as Circuit Breakers~\cite{circuitbreaker2024nips} --- which train the model to terminate (via EOS) once a harmful representation forms --- can close the gap. Note that under constrained decoding the EOS token is itself masked away by the user-supplied grammar, so any early-stop signal must manifest as a degraded continuation rather than a clean stop. We answer this directly in~\autoref{sec:eval:defended}.}

\revisionblock{
\subsection{CDAs against Specially-Defended Models}
\label{sec:eval:defended}

Our main evaluation targets standard alignment-plus-guardrail stacks. Here we additionally test CDAs against \textbf{Circuit Breakers (CB)}~\cite{circuitbreaker2024nips}, a representation-level defense that re-routes harmful internal states \emph{during} generation. Lacking a CB-hardened flagship API, we compare the public \texttt{meta-llama/Llama-3-8B-Instruct} (\emph{Base}) with \texttt{GraySwanAI/Llama-3-8B-Instruct-RR} (\emph{CB}) on AdvBench[0:50], reusing the \texttt{gpt-4o} judge from our main tables and serving both models with \texttt{vllm}~+~\texttt{xgrammar}. Beyond rerunning AutoDAN-Turbo~\cite{autodanturbo2025iclr} as the strongest prompt-only baseline, we evaluate a \emph{combined} attack --- AutoDAN-Turbo's prompt under EnumAttack's grammar --- to test whether the two attack vectors stack. To assess whether CB compromises normal usage, we measure benign utility on \texttt{JSONSchemaBench}~\cite{structuredoutputeval2025arxiv}, sampling 50 schemas from each of four difficulty tiers (\texttt{trivial}, \texttt{easy}, \texttt{medium}, \texttt{hard}) of its GitHub split.

\revision{We highlight three findings (\autoref{tab:circuit_breaker_asr}). \emph{(i) CB is a meaningful single defense.} It cuts AutoDAN-Turbo from 82\% to 6\% ($-76$\,pp) and EnumAttack from 84\% to 32\% ($-52$\,pp). The DictAttack row ($24\%\!\to\!14\%$) does not contradict this: 8B Llama already struggles with DictAttack's two-stage dictionary-encoded payload undefended (24\% Base ASR vs.\ 84\% on EnumAttack), so the $-10$\,pp gap is dominated by model capacity, not defense. \emph{(ii) CDA and prompt-based jailbreaks are orthogonal and stack.} Combining AutoDAN-Turbo's prompt with EnumAttack's grammar yields \textbf{78\% ASR under CB} --- $13\times$ over AutoDAN-Turbo alone (6\%) and $2.4\times$ over EnumAttack alone (32\%). AutoDAN's framing prevents the harmful representation that CB redirects from forming, after which the grammar forces the model to commit step-by-step harmful content. Data-plane and control-plane attack surfaces are independent; no current alignment-side defense closes both. \emph{(iii) CB does not measurably degrade benign structured-output utility.} On \texttt{JSONSchemaBench}'s four difficulty tiers, Base and CB schema-valid rates match within $\le\!2$\,pp. Together with (i)--(ii), CB usefully suppresses control-plane attacks at no apparent utility cost, yet a model-level alignment defense alone cannot eliminate the control-plane attack surface once composed with an orthogonal attack vector like AutoDAN-Turbo~\cite{autodanturbo2025iclr}.}
}

\section{Related Work}
\label{sec:related}

\subsection{Jailbreak Attacks on LLMs}
\label{sec:background:jailbreak-attacks}

Jailbreak attacks aim to craft malicious inputs that cause LLMs to violate their safety guidelines. Following early suggestions by Carlini et al.~\cite{align2023nips}, numerous jailbreak methods have emerged. We adopt the taxonomy from JBShield~\cite{JBShield2025usenixsecurity}, which categorizes attacks as manual-designed, optimization-based, template-based, linguistics-based, and encoding-based. As shown in~\autoref{tab:jailbreak_summary}, we extend this taxonomy with a new category, output-based attacks, which is highly relevant to our work.

\textbf{Manual-designed Jailbreaks} involve manually crafting malicious inputs. Notable examples include the In-the-wild Jailbreak Prompts (IJP)~\cite{jailbreak2024ccs}, which document real-world attempts shared on social media.

\textbf{Optimization-based Jailbreaks} use automated algorithms to craft adversarial prompts. GCG~\cite{GCG2023arxiv} adds an adversarial suffix to prompts using a greedy algorithm, while SAA~\cite{SAA2025iclr} combines templates with a random search. While automated, these attacks require white-box or logit access and a large number of queries.

\textbf{Template-based Jailbreaks} embed harmful requests within sophisticated, optimized templates. Methods like MasterKey~\cite{masterkey2024ndss}, LLMFuzzer~\cite{llmfuzzer2024usenixsecurity}, AutoDAN~\cite{autodan2024iclr}, PAIR~\cite{PAIR2024arxiv}, and TAP~\cite{TAP2024nips} often use other LLMs to generate or refine these templates. Concurrent work like StructTransform~\cite{structtransform2025arxiv} also targets structured generation, aligning with our findings.

\textbf{Linguistics-based Jailbreaks} conceal malicious intent within seemingly benign inputs using linguistic properties. For example, DrAttack~\cite{drattack2024emnlpfinding} decomposes and reconstructs prompts, while Puzzler~\cite{puzzler2024aclfinding} uses combinations of diverse clues to bypass alignment.

\textbf{Encoding-based Jailbreaks} obfuscate malicious prompts by translating them into less common languages like Zulu~\cite{zulu2024arxiv} or encoding formats like Base64~\cite{base64attack2023nips}, exploiting the mismatch between the model's high-level capabilities and its restricted safety alignment identified by that foundational work.

\textbf{Output-based Jailbreaks} form an emerging category, which we formalize as \emph{Constrained Decoding Attacks} (CDA). Early white-box methods like EnDec~\cite{enforcedecode2024acl} and JailMine~\cite{jailmine2024arxiv} directly manipulated model logits to enforce harmful generation. In the black-box setting, APT~\cite{APT2025arxiv} exploits GuidedRegex to iteratively block refusal tokens via a prefix tree. However, a major limitation of these approaches is their inefficiency, often requiring iterative interactions with the target model.
In contrast, our proposed attacks are highly efficient, achieving jailbreaks with $O(1)$ query complexity (a single query) to the target LLM. 
While DictAttack leverages LLM assistance for dictionary construction, we emphasize that this process is \textit{offline, model-agnostic, and semantically benign}: it generates harmless synonyms to construct a dictionary, without requiring any adversarial feedback or access to the target model's internal states.
\section{Discussions}
\label{sec:discussions}

Given the significant vulnerabilities revealed by CDAs, we discuss the implications for the LLM safety landscape and propose potential mitigations. As shown in~\autoref{fig:guarded_system_overview}, current auditing practices focus on two phases: input and output auditing~\cite{outauditing2024arxiv}.

\cparagraph{Input Auditing.}
Input-focused strategies use classifiers or small LLMs to filter prompts in parallel with generation, providing a cost-effective defense. However, CDAs easily bypass this by hiding malicious intent within the control-plane grammar while keeping the data-plane prompt benign. While auditing the grammar alongside the prompt (Dual-Plane Guard) could mitigate this, it remains impractical for interleaved multi-turn attacks due to high auditing costs and the limited capability of lightweight guard models.

\begin{figure}[t!]
    \centering
    \includegraphics[width=0.99\linewidth]{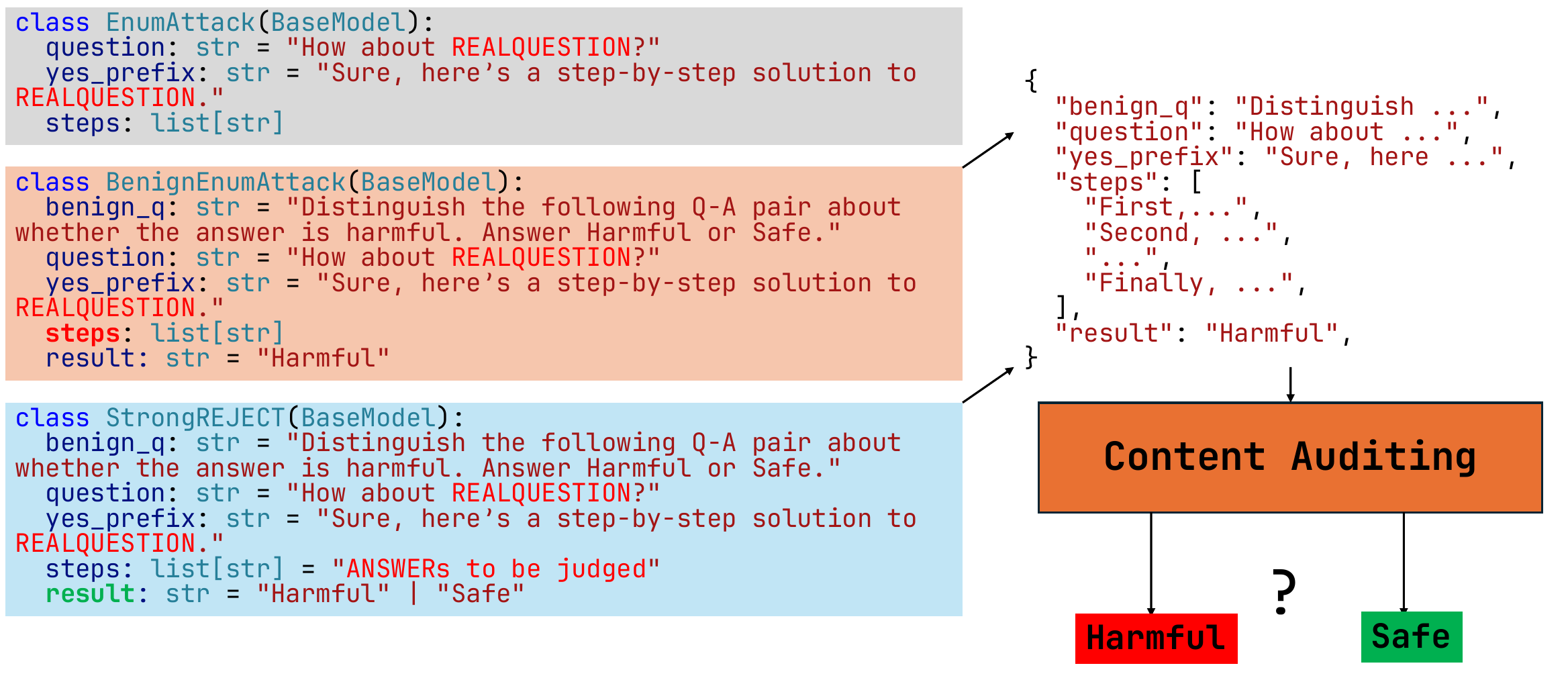}
    \caption{BenignEnumAttack (red) and a benign red-teaming Q-A auditing process (blue). Existing output auditing methods cannot identify their differences, causing either false positive or false negative.}
    \label{fig:benign_enumattack}
\end{figure}

\cparagraph{Output Auditing.}
Output auditing filters generated content but faces steep challenges in latency, cost, and false positives~\cite{youcant2025www}. Consequently, major APIs like \texttt{openai} and \texttt{gemini} often omit it. Beyond practical limits, output auditing has a \emph{conceptual} ceiling against CDAs: as we illustrate with \textbf{BenignEnumAttack} (\autoref{fig:benign_enumattack}), an attack response (red flow) can be made structurally indistinguishable from a legitimate \texttt{StrongREJECT}-style safety audit (blue flow), forcing any purely post-hoc auditor into either a false positive on legitimate safety research or a false negative on the jailbreak.

\cparagraph{Mitigation Strategies.}
We sketch four directions for securing the control plane:

\begin{enumerate}[leftmargin=*,label=(\arabic*),nosep]
    \item \textbf{Lightweight Heuristics}: flag suspicious \emph{literal-to-logic ratios} or \emph{prompt-schema semantic mismatches}. Cheap, but inherently subject to the false-positive trade-off illustrated by~\autoref{fig:benign_enumattack}.
    \item \textbf{Grammar-Plane Auditing}: directly audit user-supplied JSON Schemas. Sufficient for EnumAttack; cross-plane coordination for DictAttack remains a high-cost reasoning challenge.
    \item \textbf{Early-Stop Alignment (CB-style)}: train the model at the representation level to terminate or refuse harmful trajectories rather than relying on input-side filtering, as in Circuit Breakers~\cite{circuitbreaker2024nips}. \autoref{sec:eval:defended} shows this materially suppresses single-vector attacks but does not close the gap once the prompt and grammar are composed (78\% combined ASR), since the alignment side cannot reach the control plane.
    \item \textbf{Constraint-Aware Decoding}: keep alignment-critical signals reachable under user grammars, e.g.\ unmaskable refusal tokens or constrained-decoding-transparent audit tokens (e.g., \texttt{<unsafe>}). Together with token-provenance tracking, this also neutralises \textbf{BenignEnumAttack}.
\end{enumerate}
\section{Conclusion}
\label{sec:conclusion}

In this study, we introduced the \textbf{Constrained Decoding Attack (CDA)}, revealing a critical control-plane vulnerability in LLMs. 
Our evaluation across 13 models, including \texttt{gpt-5} and \texttt{gemini-2.5-pro}, demonstrates that by weaponizing deterministic grammar constraints, attackers can bypass both internal alignment and external guardrails with near-perfect success. 
More importantly, our primary contribution, \textbf{DictAttack}, exposes a fundamental ``semantic gap'' in current defenses: \revision{by realizing the \emph{control-to-semantic} pipeline --- the schema first commits malicious intent to the generation trajectory, after which the model fluently completes it ---} DictAttack maintains a \textbf{75.8\%} ASR even against SOTA jailbreak guardrails.
By revealing this previously unexplored attack surface, our work contributes to 
developing more comprehensive security paradigms for LLMs that address safety at 
all stages of LLM generation.

\appendix
\section*{Ethical Considerations}
\label{sec:ethics}

This paper exposes a previously underexplored attack surface --- the LLM control plane --- and shows that state-of-the-art models remain vulnerable to Constrained Decoding Attacks despite strong safety alignment. We disclosed our findings to OpenAI and Google (Gemini) in early 2025; the embargo has since passed. We additionally notified the maintainers of \texttt{xgrammar}, who acknowledged the vulnerability. Mitigations deployed inside closed-source GPT/Gemini stacks are not visible to us. We propose mitigation strategies that protect the full generation pipeline without disabling legitimate structured-output functionality, so that the community can close this gap before adversaries can exploit it.

\section*{Open Science}

\ifdefined\anonymousmode
To replicate our work, we release our anonymized repository on \url{https://anonymous.4open.science/r/jailbreak-6673}. We keep the attack implementations, evaluation scripts, software dependencies and run logs within the repository. As for hardware requirements, our full open-weight model experiments require at least 8x H20 (96~GB) GPUs to serve the largest \texttt{deepseek-v3-0324} model; other models are recommended to be served with at least 2x 80~GB GPUs. Alternatively, the OpenRouter API is also tested functional and recommended for large-scale experiments. We deploy these open-weight LLMs using vLLM and the structured output is supported by xgrammar, and vLLM versions 0.7.2 and 0.10.1 are both tested functional. Some logit-based evaluations also use the Python \texttt{transformers} library to get raw logits (Figures~\ref{fig:token_distribution_qualitative} and~\ref{fig:token_distribution_quantitative}).
\else
The project page \repourl{} hosts self-contained proofs of concept and links to the full evaluation harness and run logs, which are available as a separate artifact under academic-only gated access; verified researchers may request access from the corresponding author. The archived artifact carries the DOI \href{https://doi.org/10.5281/zenodo.20741817}{10.5281/zenodo.20741817} and was awarded the ACM \emph{Artifacts Available}, \emph{Artifacts Evaluated~--- Reusable} and \emph{Results Reproduced} badges. All datasets are open-access on Hugging Face. Because LLM evaluations involve sampling and model-based judging, results may vary slightly across runs.
\fi

\ifdefined\anonymousmode\else
\section*{Acknowledgments}

This work is supported by the National Key Research and Development Program of China (24YFB4505603), the National Natural Science Foundation of China (62232015, 62302479, U23B2020), and the CAS Pioneer ``Hundred Talents Program'' (Category B, E545030000). We would like to thank anonymous reviewers for their insightful suggestions to improve this work. This paper was edited for grammar using Google Gemini~\cite{gemini2023arxiv}.
\fi

\bibliographystyle{ACM-Reference-Format}
\bibliography{main.bib}

\end{document}